\documentclass[twocolumn]{aastex63}

\usepackage{natbib}
\usepackage{amsmath}
\usepackage{braket}
\usepackage{color}
\hypersetup{
 colorlinks=true,%
 linkcolor=black,
 citecolor=blue,
}

\begin{document}
\def\vec#1{\mbox{\boldmath $#1$}}
\newcommand{\average}[1]{\ensuremath{\langle#1\rangle} }

\title{\large{\bf Turbulent generation of magnetic switchbacks in the Alfv\'enic solar wind
}}

	\author{Munehito Shoda}
	\affiliation{National Astronomical Observatory of Japan, National Institutes of Natural Sciences, 2-21-1 Osawa, Mitaka, Tokyo, 181-8588, Japan}
	\author{Benjamin D. G. Chandran}
	\affiliation{Department of Physics and Astronomy, University of New Hampshire, Durham, NH 03824, USA}
	\author{Steven R. Cranmer}
	\affiliation{Department of Astrophysical and Planetary Sciences, Laboratory for Atmospheric and Space Physics, University of Colorado, Boulder, CO 80309, USA}

	\correspondingauthor{Munehito Shoda}
	\email{munehito.shoda@nao.ac.jp}

\begin{abstract}

  One of the most important early results from the {\it Parker Solar Probe} (PSP)
  is the ubiquitous presence of magnetic switchbacks, whose origin  is under debate.
  Using a three-dimensional direct numerical simulation of the equations of compressible magnetohydrodynamics from the corona to 40 solar radii, we investigate whether magnetic switchbacks emerge from granulation-driven Alfv\'en waves and turbulence in the solar wind.
  The simulated solar wind is an Alfv\'enic slow-solar-wind stream with a radial profile consistent with various observations, including observations from PSP.
  As a natural consequence of Alfv\'en-wave turbulence, the simulation
  reproduced magnetic switchbacks with many of the same properties as observed switchbacks, including Alfv\'enic $v$--$b$ correlation, spherical polarization (low magnetic compressibility), and a volume filling fraction that increases with radial distance.
  The analysis of propagation speed and scale length shows that the magnetic switchbacks are large-amplitude (nonlinear) Alfv\'en waves with discontinuities in the magnetic field direction. We directly compare our simulation with observations using a virtual flyby of PSP in our simulation domain.
  We conclude that at least some of the switchbacks observed by PSP are a natural consequence of the growth in amplitude of spherically polarized Alfv\'en waves as they propagate away from the Sun.
 		
\end{abstract}

\keywords{keyword for arXiv submission}

\section{Introduction} \label{sec:introduction}

Low-mass stars on the main-sequence (G, K and M dwarfs) are known to exhibit an intrinsic magnetic field \citep{Saar001,Reine09,Vidot14a,See0019b} that gives rise to a variety of magnetic activity, including coronal heating \citep{Pizzo03,Ribas05,Wrigh16,Magau20,Takas20}, flares \citep{Maeha12,Cande14,Daven16,Notsu19,Namek20}, coronal mass ejections \citep{Cranm17a,Mosch19,Argir19,Maeha20}, and stellar winds \citep{Wood005,Wood014}.
The long-term evolution of stars and planets is strongly affected by this activity:
XUV (X-ray + EUV) emission from quiescent coronae and transient flares promote the photo-evaporation of planetary atmospheres \citep{SanzF11,Johns19,Airap20},
while stellar wind can suppress the evaporative loss of the atmosphere \citep{Vidot20}.
Stellar angular momentum is extracted by magnetized stellar winds \citep{Weber67,Sakur85,Kawal88,Revil15} and/or coronal mass ejections \citep{Aarni12,Jardi20}, which results in stellar spin-down \citep{Skuma72,Barne03,Irwin09,Galle13,Galle15,Matt015}.
In fact, the solar wind is observed to transport a significant amount of angular momentum away from the Sun \citep{Finle19c,Finle20b,Finle20a}.
Since only limited and indirect observations are available for stellar winds from other low-mass stars \citep{Wood004,Kisly14,Wood014,Vidot17,Jardi19},
theoretical extrapolations from the solar-wind case are often used to infer stellar-wind properties \citep{Cranm11,Suzuk13,Suzuk18,Shoda20}.
Further understanding of the solar wind's formation is becoming increasingly important as a benchmark for stellar-wind modeling.

The classical idea of the solar wind is based on pressure-driven acceleration that leads to a transonic outflow \citep{Parke58,Parke65,Velli94}.
Early observation of supersonic plasma velocities in interplanetary space \citep{Neuge62,Neuge66} supported this idea.
This thermally-driven wind model is, however, incomplete in that it cannot reproduce the well-established anti-correlation between solar-wind velocity and coronal (freezing-in) temperature \citep{Geiss95,Steig10} nor the large wind velocities measured in fast-solar-wind streams near earth \citep{Durne72}.
In addition, the solar-wind mass flux remains nearly constant regardless of wind speed or solar activity \citep{Golds96,Wang098a,Cohen11,Cranm17a} in contrast to the thermally-driven model that predicts a sensitive dependence of the mass flux to the coronal temperature \citep{Parke65,Lamer99,Fionn18}.
To explain these observations, as well as the mass and energy budget across the transition region, a self-consistent description of coronal heating and wind acceleration via magnetic field needs to be considered \citep{Hamme82,Withb88,Hanst95,Hanst12}.
Two different types of magnetically driven solar-wind models have been proposed: wave/turbulence-driven (WTD) models and reconnection/loop-opening (RLO) models \citep[see e.g.][]{Cranm09b}.

In wave/turbulence-driven models, the solar wind is heated and accelerated by Alfv\'en waves and turbulence.
Alfv\'en waves are thought to undergo an energy cascade as a consequence of reflection-driven turbulence \citep{Velli89,Matth99,Dmitr02,Cranm05,Verdi07,Howes13,Perez13}, phase mixing \citep{Heyva83,Magya17,Magya21}, and/or parametric decay 
\citep{Sagde69,Golds78,Derby78,DelZa01,Tener13,Chand18,Revil18,Shoda18d}, which results in the observed broadband energy spectrum extending over 
orders of magnitude in length and time scales
\citep{Colem68,Belch71a,Podes07,Chen020}.
An energy cascade is indeed observed in the solar wind, and the magnitude of the resulting turbulent heating is found to be comparable to what is needed to explain the measured radial profiles of the solar-wind temperature.
 \citep{Sorri07,MacBr08,Carbo09,Baner16, Hadid17}.
Alfv\'en waves also accelerate plasma via the Alfv\'en-wave-pressure (ponderomotive) force \citep{Dewar70,Alazr71,Belch71b,Jacqu77}.
Solar-wind models with Alfv\'en-wave heating and acceleration are found to self-consistently explain the fast solar wind \citep{Hollw86,Suzuk05,Cranm07,Verdi10,Matsu12,Lione14,Shoda18a,Sakau20,Matsu21}, 
although how turbulence evolves in the solar wind is still under debate \citep{Balle16,Balle17,Zank017,Adhik19,Chand19,Tello19}.
In addition, the amplitudes of Alfv\'en waves in the solar atmosphere appear to be large enough to power the solar wind \citep{DePon07a,Baner09,McInt11,Hahn013,Sriva17}. WTD models are also able to reproduce slow solar wind when the super-radial expansion factor of the coronal magnetic field is large \citep{Ofman98,Suzuk06a,Cranm07}.
The global structure of the heliosphere can also be reproduced by WTD models \citep{Usman11,Holst14,Usman18,Revil20a}.

Compositional analysis of the solar wind indicates that (a part of) the slow solar wind may have a different origin than the fast solar wind. The
first-ionization-potential (FIP)
bias, the degree of relative enhancement of low first-ionization-potential elements, is observed to be large and variable in the slow solar wind \citep{Steig00,Stans20a}.
A large FIP bias is also observed in closed-loop regions   such as helmet streamers \citep{Raymo97,Feldm98} and active regions \citep{Widin01,Brook11,Brook15,Baker18,Dosch19}, which suggests that 
slow-solar-wind streams with large FIP bias are formed by the leakage of closed-loop plasma. 
Active-region outflows \citep{Sakao07,Harra08,Brook11,Brook15} or streamer blobs \citep{Sheel97,Wang098b,Viall15} are possible observed signatures of the leakage of the closed-field plasma.
Therefore, magnetic reconnection and the resultant opening of loops are possibly central to origin of the slow solar wind.
Models of this category are called reconnection/loop-opening (RLO) models \citep{Fisk099,Fisk003,Antio11,Higgi17,Revil20b,Wang020}.
RLO models may be particularly relevant to active stars, for which reconnection and flares occur more frequently \citep{Cande14}.

Winds from coronal holes are believed to be driven by waves and turbulence, because the energy released by magnetic reconnection in coronal holes is likely insufficient to accelerate the solar wind \citep{Cranm10,Lione16}. This conclusion, however, is challenged by the {\it Parker Solar Probe}'s recent observations of large numbers of ``magnetic switchbacks'',
which are defined as abrupt, large-angle rotations of the magnetic field that yield local reversals in the radial component of the magnetic field,
in solar wind emanating from a low-latitude coronal hole \citep{Bale019,Kaspe19}.


{\it Parker Solar Probe} (hereafter PSP) \citep{Fox0016} is a mission to observe the near-Sun solar wind by measuring electromagnetic fields \citep[FIELDS,][]{Bale016}, the distribution functions of thermal electrons, alpha particles, and protons \citep[SWEAP, ][]{Kaspe16}, high-energy particles \citep[IS$\odot$IS,][]{McCom16} and scattered white light \citep[WISPR,][]{Vourl16}.
One of the most important early results from PSP is the ubiquitous presence of sudden local magnetic polarity reversal events called magnetic switchbacks \citep{Bale019}.
The presence of switchbacks was first reported in the high-latitude wind \citep{Balog99,Matte14},
and later was also found in equatorial wind at 1 au \citep{Gosli09} and $0.3$ au \citep{Horbu18}.
The detailed observational properties are summarized in Section \ref{sec:magnetic_switchback}.

Possible scenarios for the origin of switchbacks include magnetic transient events (reconnection and/or jets) in the solar atmosphere \citep{Yamau04a,Rober18,He00020,Tener20,Sterl20,Zank020}, shear-wave interaction \citep{Landi06,Ruffo20,Shi0020} and large-amplitude turbulence \citep{Squir20}.
If magnetic transient events are the principal source of switchbacks, then this would imply that RLO-like events play a significant role even in the coronal-hole wind,
which would pose a challenge to the conventional understanding that waves and turbulence dominate the heating and acceleration in coronal-hole wind.

Motivated by the above background,
in this work we aim to investigate the theoretical properties of magnetic switchbacks in a wave/turbulence-driven solar-wind model.
For this purpose, we perform a three-dimensional direct numerical simulation of the turbulence-driven solar wind from the coronal base to $40R_\odot$, thus directly connecting coronal dynamics and solar-wind fluctuations.

\section{Model} \label{sec:model}

\subsection{Basic equations}
We solve the three-dimensional magnetohydrodynamic equations with gravity and heating \citep[e.g.][]{Pries14}:
\begin{align}
    &\frac{\partial \rho}{\partial t} + \nabla \cdot \left( \rho \vec{v} \right) = 0, \label{eq:MHD_mass} \\
    &\frac{\partial \vec{v}}{\partial t} + \left( \vec{v} \cdot \nabla \right) \vec{v} = - \frac{1}{\rho} \nabla p + \frac{1}{4 \pi \rho} \left( \nabla \times \vec{B} \right) \times \vec{B} + \vec{g}, \label{eq:MHD_eom} \\
    &\frac{\partial \vec{B}}{\partial t} - \nabla \times \left( \vec{v} \times \vec{B} \right) = 0, \label{eq:MHD_induction} \\
    &\frac{\partial e_{\rm int}}{\partial t} + \nabla \cdot \left( e_{\rm int} \vec{v} \right) + p \nabla \cdot \vec{v} = L, \label{eq:MHD_energy}
\end{align}
where $\rho$ is the mass density, $p$ is the pressure, $\vec{v}$ is the velocity field, $\vec{g}$ is the gravitational acceleration, $\vec{B}$ is the magnetic field, $e_{\rm int}$ is the internal energy per unit volume and $L$ is the heat injection per unit volume and time, respectively.
The internal energy and pressure are related by
\begin{align}
    e_{\rm int} = \frac{p}{\gamma-1},
\end{align}
where $\gamma = 5/3$ is the specific heat ratio of adiabatic gas.
Mathematically, Eq. (\ref{eq:MHD_induction}) ensures that $\nabla \cdot \vec{B} = 0$ as long as the magnetic field is divergence-free at any given time.
In numerically solving the Eq.s (\ref{eq:MHD_mass})--(\ref{eq:MHD_energy}), however, numerical error gives rise to non-zero $\nabla \cdot \vec{B}$ that could result in an unphysical solution.
To overcome this issue, a scalar variable $\zeta$ is added to the basic equations \citep{Dedne02}.
\begin{align}
    &\frac{\partial \vec{B}}{\partial t} - \nabla \times \left( \vec{v} \times \vec{B} \right) = - \nabla \zeta, \label{eq:MHD_induction_modify} \\
    &\frac{\partial}{\partial t} \zeta + c_h^2 \nabla \cdot \vec{B} =  - \frac{c_h^2}{c_p^2} \zeta. \label{eq:MHD_dedner}
\end{align}
Eq.s (\ref{eq:MHD_mass}), (\ref{eq:MHD_eom}), (\ref{eq:MHD_energy}), (\ref{eq:MHD_induction_modify}), (\ref{eq:MHD_dedner}) are the basic equations used in this work.

We solve the basic equations in the spherical coordinate system $(r, \theta, \phi)$, in which the nabla operator is given by
\begin{align}
    \nabla = \vec{e}_r \frac{\partial}{\partial r} + \vec{e}_\theta \frac{1}{r} \frac{\partial}{\partial \theta}+ \vec{e}_\phi \frac{1}{r \sin \theta} \frac{\partial}{\partial \phi}, \label{eq:nabla_spherical}
\end{align}
where $\vec{e}_{r,\theta,\phi}$ denotes the unit vector in $r$, $\theta$, $\phi$ direction.
The extension of the simulation domain in the $\theta$ direction is only $4.32 \times 10^{-2} {\rm \ rad}$, and thus, we can assume, without loss of generality, that $\theta \approx \pi/2$ or equivalently $\sin \theta \approx 1$.
The $\nabla$ operator in this local spherical coordinate system is given by
\begin{align}
    \nabla = \vec{e}_r \frac{\partial}{\partial r} + \vec{e}_\theta \frac{1}{r} \frac{\partial}{\partial \theta}+ \vec{e}_\phi \frac{1}{r} \frac{\partial}{\partial \phi}. \label{eq:nabla}
\end{align}
Using Eq. (\ref{eq:nabla}) and after some algebra, the basic equations (\ref{eq:MHD_mass}), (\ref{eq:MHD_eom}), (\ref{eq:MHD_energy}), (\ref{eq:MHD_induction_modify}), (\ref{eq:MHD_dedner}) are given in the form of a conservation law,
\begin{align}
    \frac{\partial}{\partial t} \vec{U} + \frac{1}{r^2} \frac{\partial}{\partial r} \left( r^2 \vec{F}_r \right) + \frac{1}{r} \frac{\partial}{\partial \theta} \vec{F}_\theta + \frac{1}{r} \frac{\partial}{\partial \phi} \vec{F}_\phi = \vec{S},
\end{align}
where $\vec{U}$, $\vec{F}_{r,\theta,\phi}$ and $\vec{S}$ are the conserved variables, corresponding fluxes in each direction, and source terms, respectively, which are given by
\begin{align}
    \vec{U} = \left(
    \begin{array}{c}
        \rho \\
        \rho v_r \\
        \rho v_\theta \\
        \rho v_\phi \\
        \tilde{B}_r \\
        \tilde{B}_\theta \\
        \tilde{B}_\phi \\
        e \\
        \zeta
    \end{array}
    \right), \ 
    \vec{F}_\alpha = \left(
    \begin{array}{c}
        \rho v_\alpha \\
        \rho v_\alpha v_r - \tilde{B}_\alpha \tilde{B}_r + p_T \delta_{\alpha,r} \\
        \rho v_\alpha v_\theta - \tilde{B}_\alpha \tilde{B}_\theta + p_T \delta_{\alpha,\theta} \\
        \rho v_\alpha v_\phi - \tilde{B}_\alpha \tilde{B}_\phi + p_T \delta_{\alpha,\phi} \\
        v_\alpha \tilde{B}_r - v_r \tilde{B}_\alpha + \zeta \delta_{\alpha,r} \\
        v_\alpha \tilde{B}_\theta - v_\theta \tilde{B}_\alpha + \zeta \delta_{\alpha,\theta} \\
        v_\alpha \tilde{B}_\phi - v_\phi \tilde{B}_\alpha + \zeta \delta_{\alpha,\phi} \\
        (e + p_T) v_\alpha - \tilde{B}_\alpha \left( \vec{v} \cdot \vec{\tilde{B}} \right) \\
        c_h^2 \tilde{B}_\alpha 
    \end{array}
    \right),
\end{align}
\begin{align}
    \vec{S} = \left(
    \begin{array}{c}
        0 \\
        \rho \left( v_\theta^2 + v_\phi^2 \right)/r + \left( 2p + \tilde{B}_r^2 \right)/r - \rho g \\
        \left( \tilde{B}_r \tilde{B}_\theta - \rho v_r v_\theta \right) / r \\
        \left( \tilde{B}_r \tilde{B}_\phi - \rho v_r v_\phi \right) / r \\
        \dfrac{2\zeta}{r} \\
        \left( v_r \tilde{B}_\theta - v_\theta \tilde{B}_r \right) / r \\
        \left( v_r \tilde{B}_\theta - v_\theta \tilde{B}_r \right) / r \\
        - \rho g v_r + L \\
        - \dfrac{c_h^2}{c_p^2} \zeta
    \end{array}
    \right),
\end{align}
where $\delta_{\alpha, \beta}$ is the Kronecker delta and $\tilde{\vec{B}} = \vec{B}/ \sqrt{4 \pi}$. 
$e$ and $p_T$ are the total energy per unit volume and total pressure given by
\begin{align}
    e &= e_{\rm int} + \frac{1}{2} \rho v^2 + \frac{\tilde{B}^2}{2}, \ \ \ p_T = p + \frac{\tilde{B}^2}{2}.
\end{align}
Assuming that hydrogen is ionized throughout the simulation domain, the equation of state for a fully-ionized hydrogen plasma is used to obtain the temperature:
\begin{align}
    p = 2 n_p k_B T, \ \ \  n_p = \rho/m_p,
\end{align}
where $n_p$ and $m_p$ are the proton number density and mass, respectively. 

$g$ is the absolute value of the gravitational acceleration $\vec{g}$ at radial distance $r$.
\begin{align}
    \vec{g} = - g \vec{e}_r,  \ \ \ \ g = \frac{G M_\odot}{r^2},
\end{align}
where $G$ is the gravitational constant and $M_\odot$ is the solar mass.

\begin{table*}[t!]
\centering
  \begin{tabular}{p{7em} p{7em} p{7em} p{7em} p{7em} p{7em} p{4em}} 
    \begin{tabular}{c} $r/R_\odot$  \hspace{3em} \end{tabular} & \begin{tabular}{c} $\rho$ \hspace{2em} \\ ${\rm [g \ cm^{-3}]}$ \hspace{2em} \end{tabular} & \begin{tabular}{c} $v_r$ \hspace{2em} \\ ${\rm [km \ s^{-1}}]$ \hspace{2em} \end{tabular} & \begin{tabular}{c} $\delta v_\perp$ \hspace{2em} \\ ${\rm [km \ s^{-1}]}$ \hspace{2em} \end{tabular} & \begin{tabular}{c} $\delta z^+_\perp $ \hspace{2em} \\ ${\rm [km \ s^{-1}}]$ \hspace{2em} \end{tabular} & \begin{tabular}{c} $\delta z^-_\perp$ \hspace{2em} \\ $[{\rm km \ s^{-1}}]$ \hspace{2em} \end{tabular} & \begin{tabular}{c} $\sigma_c$ \hspace{2em} \end{tabular} \rule[-3.5pt]{0pt}{16pt}   \\ \hline \hline
    \hspace{0.0em} $1.05$ & \hspace{-0.5em} $2.92 \times 10^{-16}$ & \hspace{0.75em} $0.837$ & \hspace{1.0em} $33.5$  & \hspace{1.0em} $66.6$ & \hspace{1.0em} $2.86$ & \hspace{-0.5em} $0.996$ \rule[-5.5pt]{0pt}{20pt} \\
    \hspace{0.0em} $2.00$ & \hspace{-0.5em} $2.09 \times 10^{-18}$ & \hspace{1.0em} $32.8$ & \hspace{1.0em} $88.3$  & \hspace{1.125em} $176$ & \hspace{1.0em} $5.72$ & \hspace{-0.5em} $0.998$ \rule[-5.5pt]{0pt}{20pt} \\
    \hspace{0.0em} $3.00$ & \hspace{-0.5em} $3.48 \times 10^{-19}$ & \hspace{1.0em} $87.4$ & \hspace{1.125em} $121$  & \hspace{1.125em} $243$ & \hspace{1.0em} $12.0$ & \hspace{-0.5em} $0.995$ \rule[-5.5pt]{0pt}{20pt} \\
    \hspace{0.0em} $5.00$ & \hspace{-0.5em} $6.23 \times 10^{-20}$ & \hspace{1.125em} $175$ & \hspace{1.125em} $144$  & \hspace{1.125em} $288$ & \hspace{1.0em} $26.3$ & \hspace{-0.5em} $0.984$ \rule[-5.5pt]{0pt}{20pt} \\
    \hspace{0.0em} $10.0$ & \hspace{-0.5em} $9.97 \times 10^{-21}$ & \hspace{1.125em} $273$ & \hspace{1.125em} $129$  & \hspace{1.125em} $258$ & \hspace{1.0em} $31.8$ & \hspace{-0.5em} $0.970$ \rule[-5.5pt]{0pt}{20pt} \\
    \hspace{0.0em} $20.0$ & \hspace{-0.5em} $2.05 \times 10^{-21}$ & \hspace{1.125em} $334$ & \hspace{1.0em} $92.1$  & \hspace{1.125em} $184$ & \hspace{1.0em} $22.5$ & \hspace{-0.5em} $0.971$ \rule[-5.5pt]{0pt}{20pt} \\
    \hspace{0.0em} $30.0$ & \hspace{-0.5em} $8.50 \times 10^{-22}$ & \hspace{1.125em} $356$ & \hspace{1.0em} $70.6$  & \hspace{1.125em} $141$ & \hspace{1.0em} $17.5$ & \hspace{-0.5em} $0.970$ \rule[-5.5pt]{0pt}{20pt} \\
    \hspace{0.0em} $39.5$ & \hspace{-0.5em} $4.42 \times 10^{-22}$ & \hspace{1.125em} $373$ & \hspace{1.0em} $59.4$  & \hspace{1.125em} $120$ & \hspace{1.0em} $12.0$ & \hspace{-0.5em} $0.980$ \rule[-5.5pt]{0pt}{20pt} \\ \hline
  \end{tabular}
  \vspace{1.0em}
  \caption{Summary of the averaged properties of the simulated solar wind. Columns from left to right stand for radial distance, averaged mass density, averaged radial (wind) velocity, rms perpendicular (wave/turbulence) velocity, rms outward perpendicular Els\"asser field, rms inward perpendicular Els\"asser field, and normalized cross helicity $\sigma_c = ({\delta z_\perp^+}^2 - {\delta z_\perp^-}^2)/({\delta z_\perp^+}^2 + {\delta z_\perp^-}^2)$, respectively.}
  \vspace{0.5em}
  \label{table:parameters}
\end{table*}

The energy source term $L$ is inlcuded to mimic the thermal conduction that works to homogenize the temperature profile.
To speed up the simulation, we implement exponential cooling to a given reference temperature profile by setting 
\begin{align}
    L = - \frac{1}{\tau_{\rm cnd}} \left( e_{\rm int} - e_{\rm int,ref} \right),
\end{align}
where $e_{\rm int,ref}$ is the internal energy corresponding to the reference temperature $T_{\rm ref}$,
\begin{align}
    T_{\rm ref} = T_i \xi_T + T_o \left( 1 -  \xi_T \right), \ \ 
    \xi_T = \min \left(1, \frac{1.44}{r/R_\odot} \right), 
\end{align}
where $T_i$ and $T_o$ represents the asymptotic behaviors of the reference temperature in the inner and outer sides of the simulation domain, which are given by
\begin{align}
    T_i = 10^6 {\rm \ K}, \ \ \ \ T_o = 1.2 \times 10^6 \left( r/R_\odot \right)^{-1/2} {\rm \ K}.
\end{align}
We set the relaxation time $\tau_{\rm cnd} = 1{\rm \ s}$ to ensure that the thermal relaxation is faster than the turbulent evolution. Note that this is comparable to the time scale of thermal conduction in the solar wind ($\sim 4 {\rm \ s}$, assuming a number density of $10^5 {\rm \ cm^{-3}}$, temperature of $10^6 {\rm \ K}$, and temperature scale length of $10^{5} {\rm \ km}$).

\subsection{Simulation domain and boundary conditions}

The simulation domain is a quadrangular pyramid rooted in the coronal base.
The range of the simulation domain is defined by
$1.02 \le r/R_\odot \le 40$, $-\theta_{\rm max} \le \theta \le \theta_{\rm max}$, $-\phi_{\rm max} \le \phi \le \phi_{\rm max}$,
where $\theta_{\rm max} = \phi_{\rm max} = 2.16 \times 10^{-2} {\rm \ rad}$.
The horizontal size of the simulation domain is $30,000{\rm \ km}$ at the coronal base, which is comparable to the scale of super granulation \citep{Leigh62}.
We employ $(16000,216,216)$ uniform grid points in the $(r, \theta, \phi)$ directions.

Periodic boundary conditions are imposed in the $\theta$ and $\phi$ directions.
Because the solar wind is supersonic and super-Alfv\'enic at $r/R_\odot=40$, free boundary conditions are imposed at the outer boundary.
At the inner boundary, fixed boundary conditions are imposed on the  density, radial velocity, temperature and radial magnetic field: $\rho_{\rm inner} = 4.0 \times 10^{-16} {\rm \ g \ cm^{-3}}$, $T_{\rm inner} = 1.0 \times 10^6 {\rm \ K}$, $v_{r,{\rm inner}} = 0.0 {\rm \ km \ s^{-1}}$ and $B_{r,{\rm inner}} = 1.1 {\rm \ G}$, 
where the subscript ``inner'' denotes the value at the inner boundary.
The transverse velocity and magnetic field are given in terms of Els\"asser variables $z_{\theta,\phi}^\pm$ defined by
\begin{align}
    z^\pm_{\theta,\phi} = v_{\theta,\phi} \mp \frac{B_{\theta,\phi}}{\sqrt{4 \pi \rho}}.
\end{align}
To ensure that the velocity and magnetic field are divergence-free at the inner boundary, the
Els\"asser variables are given in terms of stream functions.
\begin{align}
    z^\pm_{\theta,{\rm inner}} = \frac{1}{r} \frac{\partial}{\partial \phi} \psi^\pm_{\rm inner}, \ \ 
    z^\pm_{\phi,{\rm inner}} = - \frac{1}{r} \frac{\partial}{\partial \theta} \psi^\pm_{\rm inner},
\end{align}
where $z_{\theta,\phi,{\rm inner}}^\pm$ denote the $\theta$- and $\phi$-components of Els\"asser variables at the inner boundary and $\psi^\pm_{\rm inner}$ are the corresponding stream functions.

The stream function of the upward component $\psi^+_{\rm inner}$ is given such that it has broadband spectra in space and time:
\begin{align}
    \psi^+_{\rm inner} \propto \sum_{k_\theta = 1}^{n_\theta} & \sum_{k_\phi = 1}^{n_\phi} \sum_{f}  C \left( k_\theta,k_\phi,f \right) \nonumber \\
    \exp & \left[ 2 \pi \left( \frac{k_\theta \theta}{2 \theta_{\rm max}} + \frac{k_\phi \phi}{2 \phi_{\rm max}} +  f t \right) \right],
    \label{eq:stream_function_plus}
\end{align}
where $k_\theta$ and $k_\phi$ denote the normalized wavenumbers in the $\theta$ and $\phi$ directions, $f$ is the frequency, and $C \left( k_\theta, k_\phi, f \right)$ stands for the amplitude of the mode.

\begin{figure*}[!t]
\centering
\includegraphics[width=175mm]{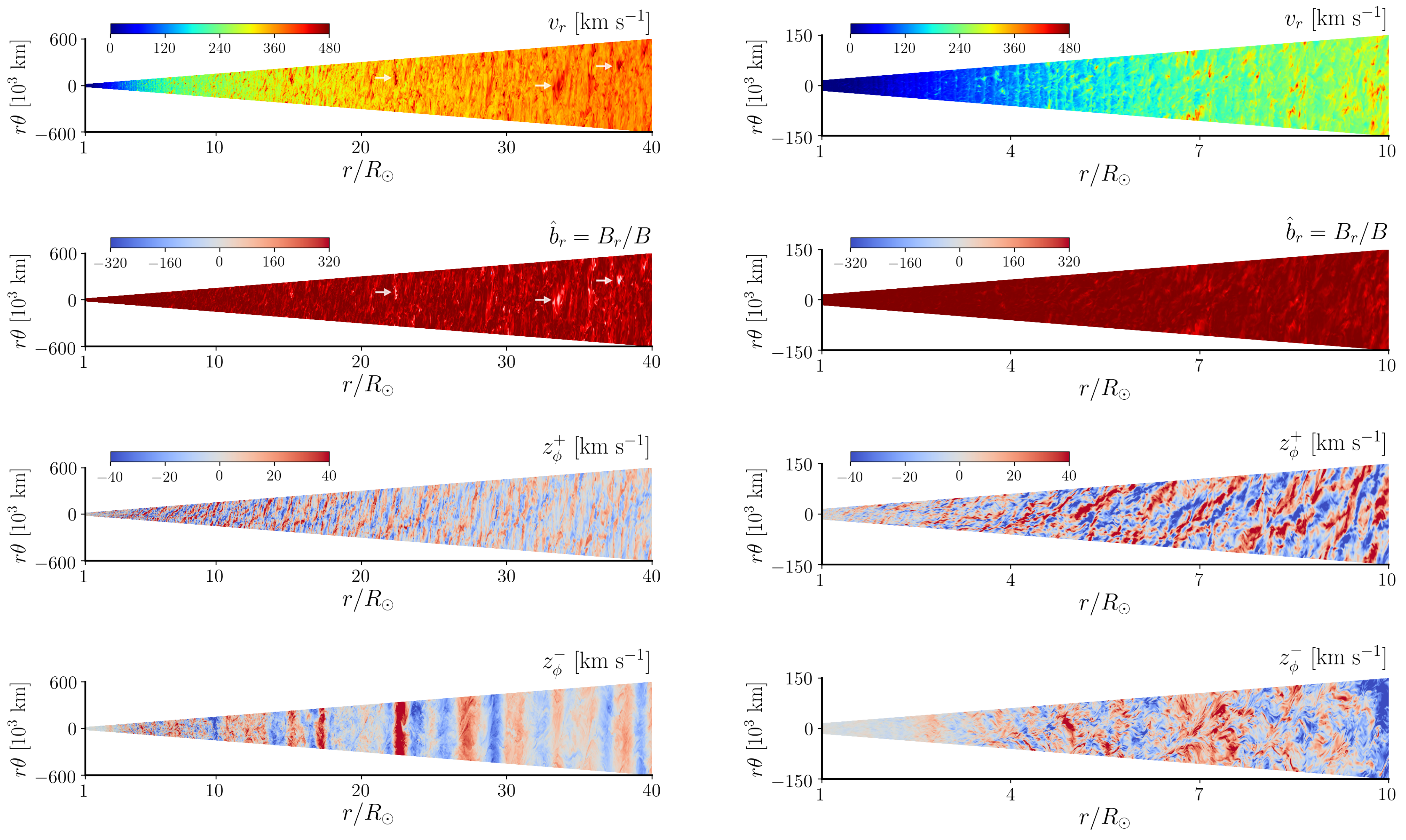}
\vspace{1em}
\caption{Simulation snapshots in the $r\theta$ plane ($\phi=-\phi_{\rm max}/3$).
Left panels cover the whole simulation domain and the right panels zoom in to the region $1<r/R_\odot<10$.
Note that the vertical axis is scaled up by $200$ times for better visualization.
Panels from top to bottom show the 
radial velocity $v_r$ in units of ${\rm km \ s^{-1}}$, 
normalized radial magnetic field $\hat{b}_r = B_r/B$,
$\phi$ component of outward Els\"asser variable $z_\phi^+$ in units of ${\rm km \ s^{-1}}$,
$\phi$ component of inward Els\"asser variable $z_\phi^-$ in units of ${\rm km \ s^{-1}}$,
respectively.
Large fluctuations in $B_r$ and associated the velocity enhancements are highlighted by white arrows. 
\added{The animation of this figure runs from simulation time, $t=0$ to $t=200$ minutes, where $t=0$ is when the system reaches a quasi-steady state.} \\
(An animation of this figure is available.)
}
\label{fig:ss_vertical_cent}
\vspace{1.5em}
\end{figure*}

Reduced MHD simulations show that the spectral index of
upward Els\"asser (Alfv\'en-wave) energy with respect to perpendicular wave number $k_\perp = \sqrt{k_\theta^2 + k_\phi^2}$ is approximately $-0.8$ at the coronal base, when the perpendicular outer scale of the turbulence in the chromosphere is sufficiently small
 \citep[see Figure 7 of][]{Chand19}.
Considering this result, the injected Alfv\'en waves are assumed to exhibit a spectral index of $-2/3$  with respect to $k_\perp$ and a broken power-law spectrum with respect to $f$. 
This is achieved by setting
\begin{align}
    C \left( k_\theta,k_\phi,f \right) = k_\perp^{-11/6}/\max \left( 1,\sqrt{f/f_{\rm inj}} \right),
\end{align}
which gives a flat frequency spectrum at $f < f_{\rm inj}$ and a $1/f$ spectrum at $f > f_{\rm inj}$.

We set $n_\theta = n_\phi = 27$ for the following reason.
To represent a broadband spectrum, a large number needs to be chosen for $n_\theta$ and $n_\phi$.
Although the maximum value for $n_\theta$ and $n_\phi$ (the Nyquist wavenumber) is 108 in our case, care needs to be taken so that the maximum wavenumber mode imposed at the inner boundary does not suffer from immediate numerical dissipation.
To avoid the numerical dissipation, any imposed wave modes need to be resolved by at least a few grid points.
For this reason, we set the maximum wavenumber of the imposed mode $n_\theta$ and $n_\phi$ to be 27, which is resolved by 8 grid points and thus is numerically non-diffusive.
$f_{\rm inj}$ corresponds to the energy-injection scale in terms of frequency.
We set $f_{\rm inj} =10^{-3} {\rm \ Hz}$, which corresponds to the timescale of the solar granulation \citep{Hirzb99}, which is believed to be the primary source of the Alfv\'en waves launched by the Sun \citep{Stein98}.

The frequency bins in Eq. (\ref{eq:stream_function_plus}) are uniformly distributed in $10^{-4} {\rm \ Hz} \le f \le 10^{-2} {\rm \ Hz}$.
The amplitude of $\psi^+$ is tuned so that the rms amplitude of the velocity fluctuations of the upward Alfv\'en waves is $30 {\rm \ km \ s^{-1}}$ at the lower boundary.

At the lower boundary,
the stream function of the downward component is assumed to vanish ($\psi^-_{\rm inner} = 0$), which yields a discontinuity in the amplitude of the downward Alfv\'en waves.
However, since the amplitude of downward Alfv\'en waves is small near the bottom boundary ($\lesssim 10 {\rm \ km \ s^{-1}}$), an unphysical discontinuity is unlikely to affect the dynamics of the solar wind.

\vspace{1em}

\section{Alfv\'enic slow solar wind}

In approximately $10^5 {\rm \ s}$, the system reaches a quasi-steady state (QSS) that is not influenced by the initial conditions.
Since our interest is the dynamics of turbulence in the quasi-steady solar wind, 
we analyze the simulation data in the QSS with time duration of $\tau_{\rm dur} = 1.2 \times 10^4 {\rm \ s}$ and time cadence of $\Delta t = 6 {\rm \ s}$. 
The actual time step in the simulation is in the order of $10^6$.

In Table 1, we show the averaged properties of the simulated solar wind at several radial distances.
According to the the values shown in Table 1,
the simulated solar wind exhibits an averaged mass-loss rate of $2.4$--$2.7 \times 10^{-14} \ M_\odot {\rm \ yr^{-1}}$ \citep[which is in the observed range of $2$--$3 \times 10^{-14} \ M_\odot {\rm \ yr^{-1}}$, see Figure 1 of][]{Cranm17b} and
is characterized by its relatively low speed $v_r \lesssim 400 {\rm \ km \ s^{-1}}$ and high normalized cross helicity $\sigma_c \gtrsim 0.95$, and thus is categorized as Alfv\'enic slow solar wind \citep{Marsc81,DAmic15,Damic19}.
Since the first perihelion of PSP is dominated by Alfv\'enic slow solar wind \citep{Bale019}, it is worthwhile to compare our simulation results with the encounter-1 observations.

\subsection{Overview on $r\theta$-plane}

\begin{figure}[t]
\begin{flushleft}
\ \ \ \ \ \includegraphics[width=76mm]{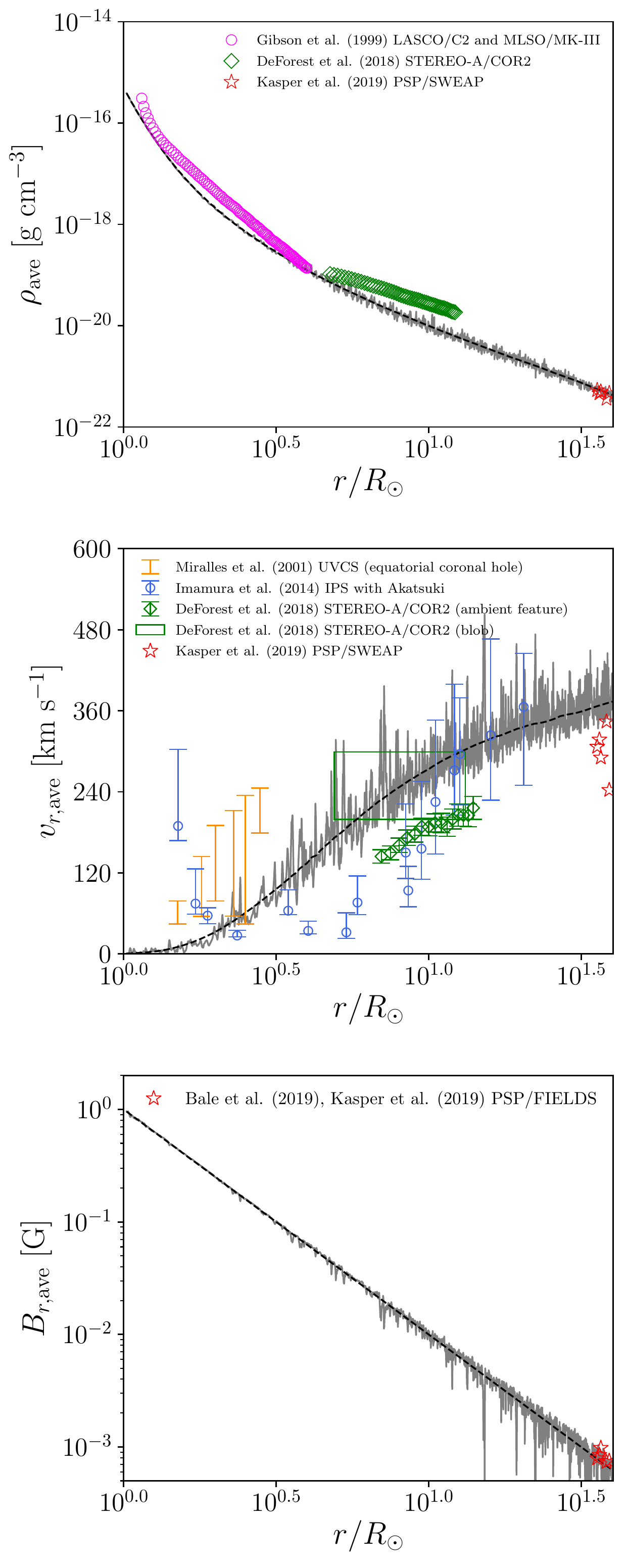}
\end{flushleft}
\vspace{-1em}
\caption{Radial profiles of the mass density (top), radial velocity (middle) and radial magnetic field (bottom).
Black dashed and grey solid lines represent averaged and non-averaged (snapshot) profiles, respectively.
Symbols show the observed data from \citet{Gibso99} (pink circles),
\citet{Defor18} (green diamonds and rectangle), \citet{Kaspe19} (red stars), \citet{Miral01} (orange bars), \citet{Imamu14} (blue circles).
}
\label{fig:mean_field}
\end{figure}

To see the radial evolution of the solar wind and turbulence therein, 
visualization of the vertical ($r \theta$) plane is useful.
Figure \ref{fig:ss_vertical_cent} shows a snapshot of the QSS in the $r \theta$ plane. 
Left and right panels visualize the whole simulation domain ($1 \le r/R_\odot\le 40$) and the near-Sun region ($1 \le r/R_\odot\le 10$), respectively.

The panels that display the outward and inward Els\"asser variables show the complex nature of turbulent transport in the solar wind.
Comparing the spatial structure of $z_\phi^+$ and $z_\phi^-$ at $1<r/R_\odot<10$,
one finds finer structures in $z_\phi^-$.
This structure difference was previously found in a numerical simulation of fast solar wind \citep{Shoda19}.
Farther from the sun where the solar wind is super Alfv\'enic, 
the chaotic nature of $z_\phi^-$ disappears, and instead large-scale structure with respect to $\theta$ is evident.

Local reversals of the radial magnetic field are highlighted by arrows in the map of $B_r/B$.
In the near-Sun solar wind, 
the radial magnetic field is found to exhibit small fluctuations.
Beyond $r \sim 25R_\odot$, 
$B_r$ occasionally locally reverses to form magnetic switchbacks \citep{Bale019}.
Figure \ref{fig:ss_vertical_cent} also shows that large fluctuations in $B_r$ are associated with local enhancements in $v_r$, which are also highlighted by arrows.
These switchbacks are discussed in more detail in Section \ref{sec:magnetic_switchback}.

\subsection{Radial profiles of the mean field}

To analyze the large-scale properties of the simulated solar wind, we average the plasma properties and magnetic field over the $\theta\phi$ plane and time.
For any given variable $X$, 
its horizontal average is defined by
\begin{align}
    \overline{X} \equiv \frac{1}{4 \theta_{\rm max} \phi_{\rm max}}
    \int_{-\theta_{\rm max}}^{\theta_{\rm max}} d \theta
    \int_{-\phi_{\rm max}}^{\phi_{\rm max}} d \phi \ X, \label{eq:horizontal_average}
\end{align}
while its time average is by
\begin{align}
    \langle X \rangle &\equiv \frac{1}{\tau_{\rm dur}} \int_{0}^{\tau_{\rm dur}} d t \ X,
\end{align}
where $t = 0$ is the time when the system reached QSS.
The averaging time $\tau_{\rm dur}$ is set to $1.2 \times 10^4 {\rm \ s}$.
The average with respect to time and horizontal space is denoted by
\begin{align}
    X_{\rm ave} \equiv \overline{\langle X \rangle}.
\end{align}
The root-mean-square fluctuation amplitude of $X$, $\delta X_{\rm rms}$, is defined by
\begin{align}
    \delta X_{\rm rms} \equiv \sqrt{ \overline{ \langle X^2 \rangle} - \overline{ \langle X \rangle}^2}. \label{eq:deltaX_ave_definition}
\end{align}

The black dashed lines in Figure \ref{fig:mean_field} show the radial profiles of the averaged density $\rho_{\rm ave}$, radial velocity $v_{r,{\rm ave}}$, and radial magnetic field $B_{r,{\rm ave}}$.
Also shown in the grey solid lines are the corresponding non-averaged (snapshot) profiles at the center of the simulation domain ($\theta = \phi = 0$).
Observations from SOHO/LASCO \citep{Bruec95}, SOHO/UVCS \citep{Kohl095,Kohl097}, MLSO/MK-III, STEREO-A/SECCHI/COR2 \citep{Howar08}, Akatsuki \citep{Nakam11}, PSP/FIELDS \citep{Bale016}, PSP/SWEAP \citep{Kaspe16} are also plotted (see caption for corresponding papers).
All PSP data are retrieved from \citet{Kaspe19} as the most probable values at given radial distance in PSP's encounter~1.
In converting the proton/electron number density to the mass density, we assumed that the helium-to-proton ratio is $5\%$ in number. 
We note that, although the typical slow solar wind is found to exhibit smaller helium abundance \citep[$\sim 2.5\%$,][]{Kaspe07}, highly Alfv\'enic slow solar wind can be helium-rich with abundance as large as $5 \%$ \citep{Huang20}.

Although there are broad similarities between our simulation and the observations, there are also several differences. In the density plot, the
measurements from \citet{Defor18} systematically lie above our model possibly because they estimated the density of blobs while our model shows the density of the ambient solar wind.  
A systematic gap also exists between our model and the velocity measurements from \citet{Miral01}.
A possible reason for this discrepancy is that they reported the outflow velocity of O VI which is often found to be systematically larger than the bulk speed of the solar wind \citep{Kohl098}.

\begin{figure}[t]
\begin{flushleft}
\includegraphics[width=75mm]{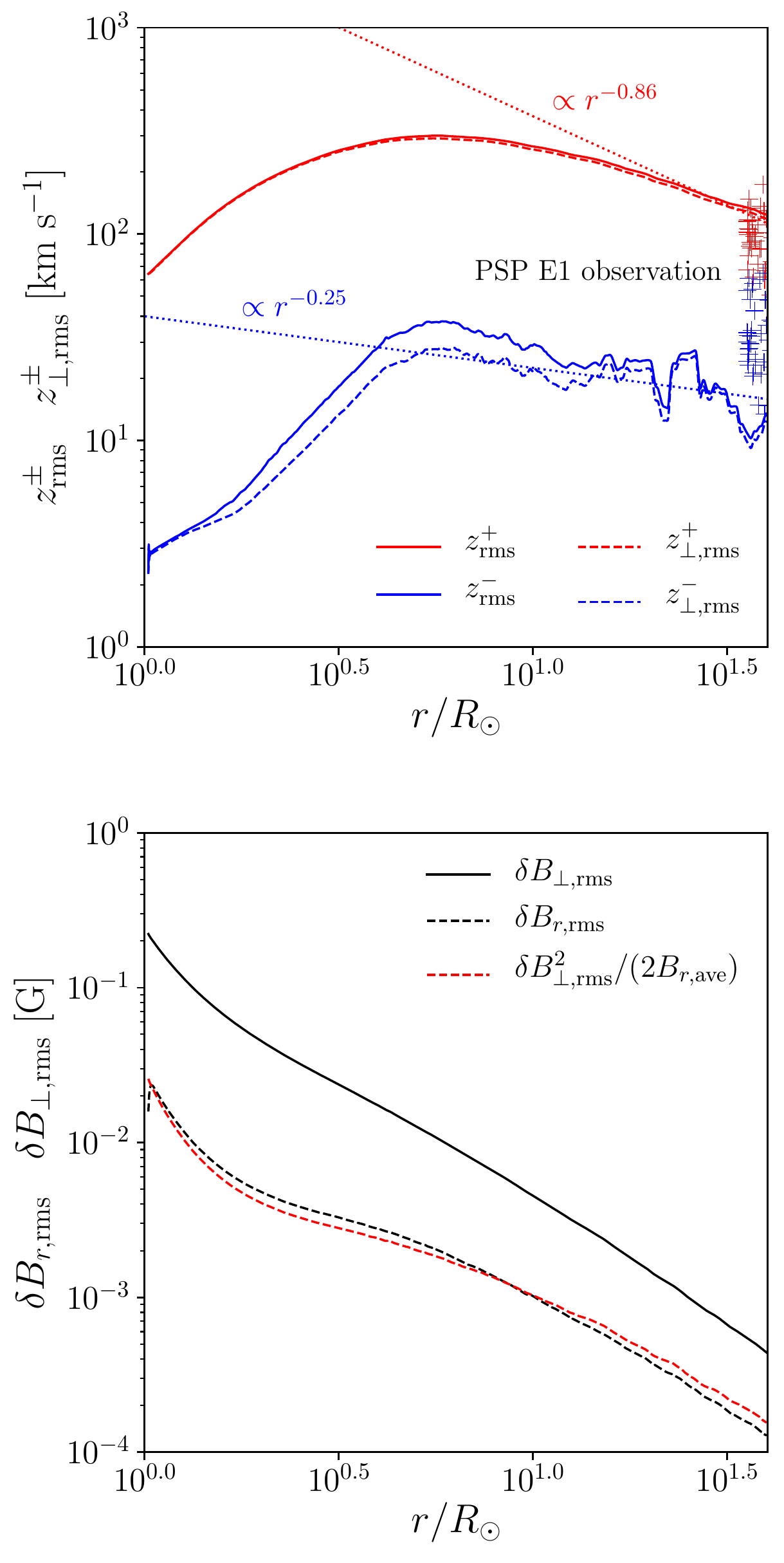}
\end{flushleft}
\vspace{-1em}
\caption{
Radial profiles of Els\"asser fields and magnetic field.
(top) The red and blue lines in the top panel show the rms amplitudes of outward ($z^+$) and inward ($z^-$) Els\"asser fields.
Dashed lines are from transverse ($\theta$ and $\phi$) components only while solid lines are from the full three components.
Crosses represent the observed amplitudes of Els\"asser fields during PSP encounter 1 \citep[retrieved from][]{Chen020}. The asymptotic radial scaling laws are indicated by dotted lines \citep{Chen020}.
(bottom) Black solid and dashed lines show the rms amplitudes of fluctuations in transverse and radial magnetic field. 
Also shown by the red dashed line is $\delta B_{\perp,{\rm rms}}^2/(2 B_{r,{\rm ave}})$, a theoretical expectation of $\delta B_{r,{\rm rms}}$ from \cite{Vasqu98}.
}
\label{fig:vb_fluctuation}
\vspace{1em}
\end{figure}

\begin{figure}[t]
\begin{flushleft}
\includegraphics[width=75mm]{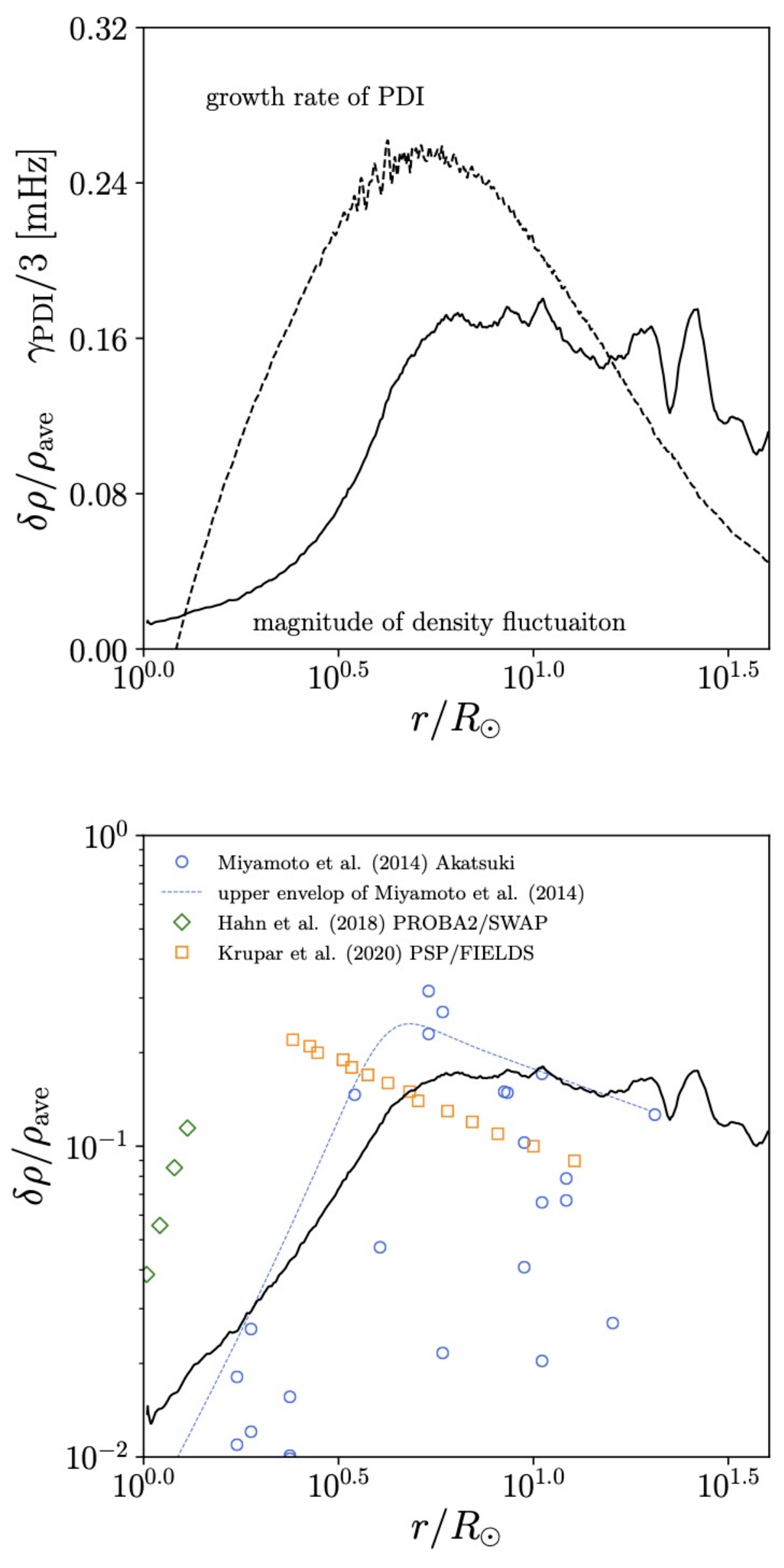}
\end{flushleft}
\vspace{-1em}
\caption{(top) Radial profiles of the fractional density fluctuation $\delta \rho_{\rm rms} / \rho_{\rm ave}$ (solid line) and the growth rate of the parametric decay instability (dashed line).
(bottom) Comparison of fractional density fluctuations from this model (solid line) and observations (symbols). 
Symbols represent the observed density fluctuation by Akatsuki \citep[blue circles,][]{Miyam14}, by PROBA2/SWAP \citep[green diamonds,][]{Hahn018}, and by PSP/FIELDS \citep[orange squares,][]{Krupa20}.
}
\label{fig:density_fluctuation}
\vspace{1em}
\end{figure}

The radial magnetic field on average scales as $B_r \propto r^{-2}$.
This is because the mean field is forced to be aligned with the simulation domain that expands radially in the spherical coordinate system.
In the snapshot data (grey line), one finds multiple sudden decreases of $B_r$ (magnetic switchbacks) that appear at $r/R_\odot \gtrsim 10$.

\subsection{Radial profiles of the fluctuation amplitudes}

Since the energy for the wind heating and acceleration lies in the turbulent fluctuations, 
how the amplitude of the turbulence varies in $r$ is worth investigating.
Figure \ref{fig:vb_fluctuation} shows the radial profiles of the amplitudes of the fluctuations in the 
Els\"asser fields ($\vec{z}^\pm = \vec{v} \mp \vec{B}/\sqrt{4 \pi \rho}$ ) and magnetic field.

In the top panel, solid lines show the rms amplitudes of outward and inward Els\"asser fields, and 
the dashed lines show the rms amplitudes of the transverse
components of the Els\"asser fields: $\vec{z}^\pm_\perp = z^\pm_\theta \vec{e}_\theta + z^\pm_\phi \vec{e}_\phi$.
Since the fluctuations are nearly transverse, the solid and dashed lines nearly overlap.
Crosses are PSP encounter-1 observations from \citet{Chen020}. Also shown by the dotted lines are asymptotic radial scalings of $z^\pm$ from \citet{Chen020} \citep[see also][]{Bavas00}.
High Alfv\'enicity is maintained throughout the simulation domain, possibly because the weaker Els\"asser field decays more rapidly \citep{Dobro80,Pouqu86}.

The bottom panel shows the profiles of the radial and transverse magnetic-field fluctuations.
The radial-field fluctuation is approximately one order of magnitude smaller than the transverse fluctuation, which is consistent with the transverse nature of the Alfv\'en wave.
According to \citet{Vasqu98}, the interaction between two oblique Alfv\'en waves with the same group velocity drives a second-order parallel-field fluctuation given by 
\begin{align}
    \delta B_\parallel = \frac{1}{2} \frac{B_\perp^2}{B_\parallel},
    \label{eq:vasques_1998}
\end{align}
consistent with earlier work by \cite{Barne74} on spherically polarized (or constant-$|\vec{B}|$) Alfv\'en waves. The red dashed line in the bottom panel of Figure \ref{fig:vb_fluctuation} shows a root-mean-square version of Eq.~(\ref{eq:vasques_1998}), which nicely overlaps the black dashed line, supporting the idea that the radial-field fluctuation is driven nonlinearly by oblique Alfv\'en waves. 
Eq. (\ref{eq:vasques_1998}) predicts that the radial-field fluctuation can be as large as the mean radial field when the Alfv\'en-wave fluctuation is comparable to the mean field.

In contrast to most reduced-MHD simulations \citep[e.g.][]{Dmitr03,Balle11,Perez13,Chand19}, our compressible-MHD simulation includes density fluctuations that can be compared with observations.
The presence of the density fluctuations in the solar wind is confirmed by several types of measurements \citep{Tu00094,Miyam14}, and these fluctuations may have an important impact on the dynamics of solar-wind turbulence \citep{Balle16,Balle17,Shoda19}.
Some theoretical studies have been able to reproduce density fluctuations consistent with observed fluctuations \citep{Suzuk05,Matsu12,Shoda18a,Adhik19,Adhik20}, which possibly result from the parametric decay instability (PDI) of Alfv\'en waves \citep{Tener13,Tener17b,Chand18,Revil18,Bowen18,Shoda18d}.
\citet{Shoda19} showed that the PDI plays an essential role in driving turbulence in the fast solar wind.
However, it is unclear whether the PDI is important also in the Alfv\'enic slow solar wind.
Here, we discuss the origin and role of density fluctuations, following an analysis similar to that of \citet{Shoda19}.

Figure \ref{fig:density_fluctuation} shows the radial profile of the fractional density fluctuation $\delta \rho_{\rm rms} / \rho_{\rm ave}$ in our model (black solid lines).
The growth rate of the parametric decay instability $\gamma_{\rm PDI}$ is also shown in the top panel (dashed line).
In calculating $\gamma_{\rm PDI}$ we numerically solved the Goldstein--Derby dispersion relation \citep{Golds78,Derby78} with frequency $f = f_{\rm inj} = 10^{-3} {\rm \ Hz}$ and added the suppression terms from wind acceleration and expansion \citep{Tener13,Shoda18d}.
Comparing the growth rate of the PDI and the magnitude of density fluctuations, one finds that the density fluctuations grow rapidly where the growth rate of the PDI is large.
This spatial correlation supports the idea of density-fluctuation generation by the PDI, consistent with what is found in fast-wind simulations \citep{Shoda18d,Shoda19}.

The bottom panel of Figure \ref{fig:density_fluctuation} compares the root-mean-square amplitudes of the simulated and observed density fluctuations.
The observed values are from Akatuski \citep{Nakam11}, PROBA2/SWAP \citep{Bergh06,Seato13} and PSP/FIELDS \citep{Bale016} (see caption for the corresponding papers).
Radio scintillation observations from \citet{Miyam14} exhibit a large scatter.
To take into account the underestimation of the local density-fluctuation amplitude by the positive-negative cancellation along the line of sight,
we focus on the upper envelope of the measurements (the blue dashed line) rather than each value (blue circles).
The upper envelope is globally consistent with our model.
EUV observations (green diamonds) are systematically higher possibly because we ignore the slow magnetosonic waves from below the transition region \citep{DeFor98,Ofman99,Kiddi12} and also the large cross-field variation in the background density along different magnetic flux tubes that has been inferred from Comet-Lovejoy observations \citep{Raymo14}.

Type-III radio burst observations (orange squares) show a similar magnitude but a somewhat different trend at $r/R_\odot<5$.
In spite of these differences, overall, the observations are broadly consistent with our model.
We note, however, that MHD simulations may overestimate the density fluctuations because they neglect collisionless damping (Barnes 1966; but see Schekhochihin et al 2016, Meyrand et al 2019), and thus care needs to be taken in the interpretation.
We also note that fast-mode waves \citep{Cranm12a} and co-rotating interaction regions \citep{Cranm13} may also contribute to the observed density fluctuations.

\vspace{1em}

\begin{table*}[t!]
\centering
  \begin{tabular}{p{17.4em} p{5.0em} p{5.0em} p{5.0em} p{5.0em} p{5.0em} p{5.8em}}
    \begin{tabular}{l} \hspace{-3em} data \end{tabular} 
    & \begin{tabular}{c} $B_{\rm ave}$ \hspace{-1.5em} \\ ${\rm [nT]}$ \hspace{-1.5em} \end{tabular}
    & \begin{tabular}{c} $B_{r,{\rm ave}}$ \hspace{-1.5em} \\ ${\rm [nT]}$ \hspace{-1.5em} \end{tabular} 
    & \begin{tabular}{c} $\left| \delta \vec{B}_{\rm rms} \right|$ \hspace{-1.5em} \\ ${\rm [nT]}$ \hspace{-1.5em} \end{tabular} 
    & \begin{tabular}{c} $\delta B_{r,{\rm rms}}$ \hspace{-1.5em} \\ ${\rm [nT]}$ \hspace{-1.5em} \end{tabular}
    & \begin{tabular}{c} $\dfrac{\delta B_{r,{\rm rms}}}{\left| \delta \vec{B}_{\rm rms} \right|}$ \hspace{-1.5em} \end{tabular}
    & \begin{tabular}{c} $\left| \dfrac{\delta \vec{B}_{\rm rms}}{B_{r,{\rm ave}}} \right|$ \hspace{-2.5em} \end{tabular}
    \rule[-5.5pt]{0pt}{20pt} \\ \hline \hline
    \hspace{-0.5em} observation (12:40--16:00 on 11/05/2018)
    & \hspace{1.9em} $81.63$
    & \hspace{1.6em} $-48.81$ 
    & \hspace{2.5em} $62.50$  
    & \hspace{2.5em} $35.76$
    & \hspace{2.3em} $0.572$
    & \hspace{2.9em} $1.280$ \rule[-5.5pt]{0pt}{20pt} \\
    \hspace{-0.5em} observation (6:20--9:40 on 11/06/2018)
    & \hspace{1.9em} $86.55$
    & \hspace{1.6em} $-71.55$ 
    & \hspace{2.5em} $47.91$  
    & \hspace{2.5em} $17.44$
    & \hspace{2.3em} $0.364$
    & \hspace{2.9em} $0.670$ \rule[-5.5pt]{0pt}{20pt} \\
    \hspace{-0.5em} simulation
    & \hspace{1.9em} $97.20$
    & \hspace{1.6em} $-79.74$ 
    & \hspace{2.5em} $55.42$  
    & \hspace{2.5em} $15.38$
    & \hspace{2.3em} $0.277$
    & \hspace{2.9em} $0.695$ \rule[-5.5pt]{0pt}{20pt} \\ \hline 
  \end{tabular}
  \vspace{1.0em}
  \caption{This table shows magnetic-field properties in the two observational periods in the encounter 1 and the corresponding simulated data displayed in Figure \ref{fig:psp_24s_average_simulation_comparison_three_column}. Note that the averaged and root-mean-squared values are defined in terms of time average over $200 {\rm \ min}.$}
  \vspace{0em}
  \label{table:psp_obs_characteristics}
\end{table*}

\begin{figure*}[t!]
\centering
\includegraphics[width=180mm]{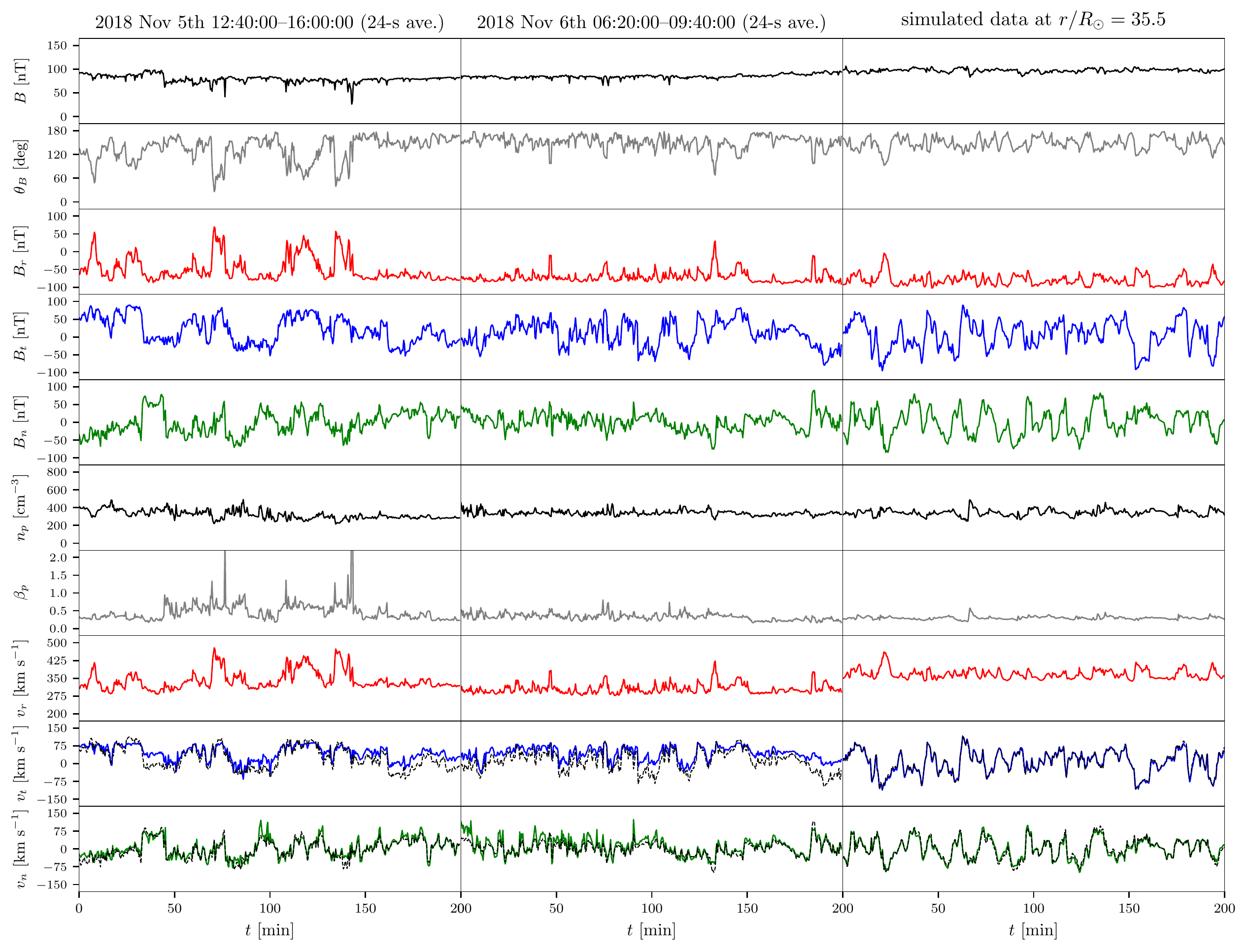}
\caption{
Comparison between PSP data averaged over 24 seconds (left and middle panels) and simulated data retrieved from a virtual PSP flyby through the simulation domain (right panels).
Two hundred minutes of data from 12:40 to 16:00 on 2018 November 5th and from 06:20 to 09:40 on 2018 November 6th are shown in the PSP data.
The top five panels show magnetic-field data (total field strength $B$, 
angle between $\vec{B}$ and the radial direction $\theta_{BR} = \arccos \left( B_r/B \right)$,
radial field $B_r$, tangential field $B_t$, and normal field $B_n$)
and the bottom five panels show plasma data (proton density $n_p$, proton beta $\beta_p$, radial velocity $v_r$, tangential velocity $v_t$, and normal velocity $v_n$).
Black dashed lines in the bottom two panels show the magnetic field in the Alfv\'en units: $B_{T,N}/\sqrt{4 \pi \rho}$.
Magnetic-field data are from MAG/FIELDS and plasma data are from SPC/SWEAP via the Coordinated Data Analysis Web (CDAWeb).
}
\label{fig:psp_24s_average_simulation_comparison_three_column}
\vspace{1em}
\end{figure*}

\section{Direct comparison with PSP data}

Since our simulation domain extends beyond the perihelions of {\it Parker Solar Probe}, a direct comparison between our simulation and PSP observations is possible.
Here we compare 
our simulation results with PSP encounter-1 data.

Around the first perihelion, PSP was nearly co-rotating with the Sun.
Although the effect of rotation is not included in our simulation, 
we simply assume that the simulation domain is co-rotating with the Sun.
The location of PSP in the simulation domain is therefore assumed to be fixed in time.
For any variable $X$, the simulated time series of encounter-1 $X_{\rm sim} (t)$ is calculated using the equation
\begin{align}
    X_{\rm sim} (t) = \left. X \left(r,\theta,\phi,t \right)\right|_{r=35.5R_\odot, \ \theta=-\theta_{\rm max}, \ \phi = - \phi_{\rm max}}.
\end{align}


Observed encounter-1 data of PSP are retrieved from the Coordinated Data Analysis Web (CDAWeb).
Magnetic fields in RTN coordinates are taken from the Level-2 data of the fluxgate magnetometer (MAG), part of the FIELDS instrument suite.
The proton density, bulk velocity and thermal velocity
are taken from the Level-3 data of the Solar Probe Cup (SPC), part of the SWEAP instrument suite, as 0th-, 1st- and 2nd-moments of the reduced distribution function. 

Two sets of two-hundred minutes of data are compared with our simulation: the interval from 2018/11/05 12:40:00 to 2018/11/05 16:00:00 and the interval from 
 2018/11/06 06:20:00 to 2018/11/06 09:40:00.
The former period corresponds 
to a relatively active phase that exhibits large fluctuations in $B_r$, and the latter 
to a relatively quiet phase that exhibits small fluctuations in $B_r$.
Table \ref{table:psp_obs_characteristics} shows the quantitative difference in the magnetic-field properties in these two periods and the simulated data. The data from Nov 5th are characterized by large $\delta B_{r,{\rm rms}}/\left| \delta \vec{B}_{\rm rms}\right|$ with little `variance anisotropy': $\delta B_{r,{\rm rms}}/\left| \delta \vec{B}_{\rm rms}\right| \approx 0.577$ and $\delta B_r \approx \delta B_t \approx \delta B_n$. On the other hand, both the data from Nov 6th and the data from our simulation exhibit much smaller values of $\delta B_{r,{\rm rms}}/\left| \delta \vec{B}_{\rm rms}\right|$.
Since our simulation captures neither high-frequency fluctuations originating from kinetic physics nor sub-grid-scale fluctuations, we take a 24-second time average of the PSP data before comparing with our simulation results. 

Figure \ref{fig:psp_24s_average_simulation_comparison_three_column} shows a direct comparison between observed (left and middle column) and simulated data (right column) of PSP.
In this comparison, the sign of the simulated magnetic field is changed ($\vec{B}_{\rm sim} \to - \vec{B}_{\rm sim}$) so that mean magnetic-field polarity is consistent with observations:
during this period PSP observed negative-polarity-dominated wind while our simulation assumes a positive-polarity-dominated wind.
A change in the sign of $\vec{B}$ does not affect the comparison because the structure of field line is invariant.
In converting bulk velocity $\vec{v}$ to proton bulk velocity $\vec{v}_p$,
we assume no differential flow between different species, thus $\vec{v} = \vec{v}_p$.
The proton number density is obtained by a simple assumption that the solar wind is composed of fully ionized hydrogen plasma.

Considering the lack of artificial fine tuning in the simulation,
the similarity between the Nov-6th observations and simulation results is surprising.
Various observational properties are also seen in the simulated time series: approximately constant density and field strength, the one-sided nature of the fluctuations in $v_r$ and $B_r$, turbulent fluctuations in $\vec{v}$ and $\vec{B}$, and the Alfv\'enic correlation between~$\vec{B}$ and~$\vec{v}$ shown in the bottom two panels of 
Figure \ref{fig:psp_24s_average_simulation_comparison_three_column}. 

Meanwhile, the Nov-5th observation exhibits several substantial differences from the simulation data.
Magnetic switchbacks (where $B_r>0$, see Section \ref{sec:magnetic_switchback} for definition) that last more than a few minutes are observed multiple times, which are not seen in the simulation data. As a result of the emergence of switchbacks, the magnetic field experiences more discontinuous jumps than the simulation predicts. 
However, as noted above and shown in
Table \ref{table:psp_obs_characteristics}, $\delta B_{r,\rm rms}/|\delta \vec{B}_{\rm rms}|$ is smaller in the simulation than in the Nov-5 data by a factor of more than~2.  Given that the fractional magnetic-field fluctuation is significantly larger in the Nov-5 data, it is perhaps not surprising that switchback features are more prominent than in the simulation. However, it remains to be seen whether a numerical simulation with larger $\delta B_{r,\rm rms}/|\delta \vec{B}_{\rm rms}|$ (and perhaps higher numerical resolution, as discussed above) could explain the data, or whether additional physical ingredients (such as impulsive forcing at the coronal base) are needed.
It remains to be seen whether a numerical simulation with higher resolution and/or amplitude could could explain the data, or whether additional physical ingredients (such as impulsive forcing at the coronal base) are needed.

\begin{figure}[t]
\begin{flushleft}
\includegraphics[width=82mm]{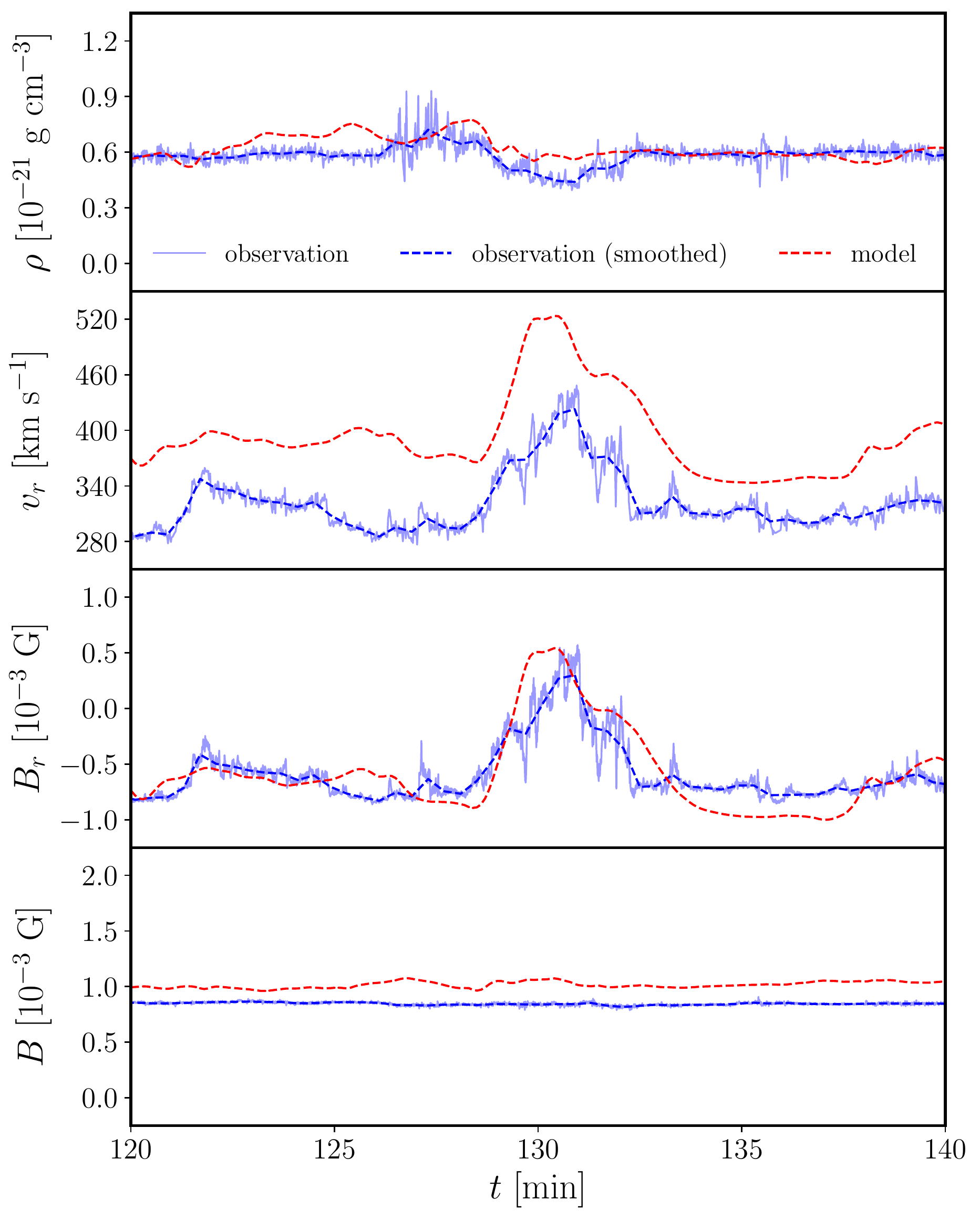}
\end{flushleft}
\caption{
Comparison between simulation results \added{(red dashed line)} and PSP observations \added{(blue lines)} for an individual switchback. The numerical data correspond to the fixed-point time series at $(r/R_\odot,\theta/\theta_{\rm max},\phi/\phi_{\rm max}) = (35, 0.815, 0.046)$ between $t=120 \ {\rm min}$ and $t = 140 \ {\rm min}$.
The PSP data are from the interval 08:22:30 to 08:42:30 on November 6, 2018, \deleted{averaged over $24 {\rm \ s}$}
\added{with (blue dashed line) and without (blue semi-transparent line) time averaging over $24 {\rm \ s}$}
.
}
\label{fig:switchback_te_35Rs}
\vspace{1em}
\end{figure}

In Figure \ref{fig:switchback_te_35Rs}, we compare a single switchback observed in our simulation at $r/R_\odot=35$
with an individual switchback observed by PSP on November 6, 2018 \citep[see ][for a similar comparison]{Zank020}. 
\added{For the PSP measurements, we show both the full data and the data averaged over 24~s, the latter of which is to remove the high-frequency fluctuations beyond the capability of our model.}
The simulated and observed switchbacks exhibit similar characteristics, including similar durations and amplitudes, consistent with the hypothesis that the switchbacks seen by PSP emerge naturally from the dynamics of solar-wind turbulence, even without impulsive forcing near the Sun. 
The detailed statistical properties of switchbacks are discussed in the next section.

\section{Magnetic switchbacks \label{sec:magnetic_switchback}}
Observationally, magnetic switchbacks are defined as regions in which the radial field is reversed against the ambient region.
Since the mean radial magnetic field is positive in our simulation, switchbacks are defined as regions in which $B_r<0$, and switchback boundaries are defined as surfaces on which $B_r=0$.
In the previous section, we have already shown the presence of switchbacks in our simulation.

In this section, after reviewing the observational properties of magnetic switchbacks, we present several detailed analyses of switchbacks, focusing on their physical properties and the comparison with observations.

\begin{figure*}[!t]
\centering
\includegraphics[width=160mm]{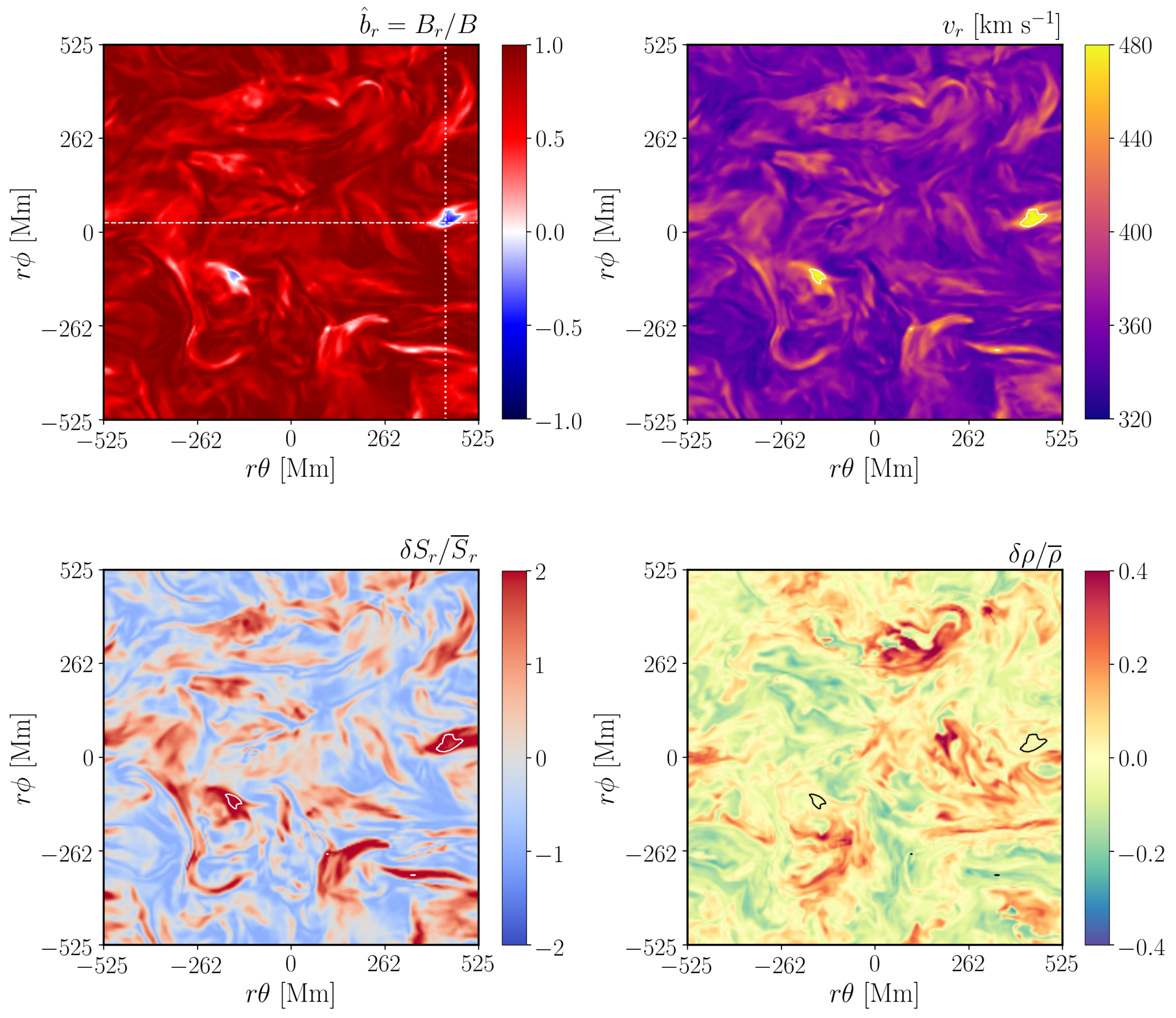}
\vspace{0em}
\caption{
Horizontal ($\theta \phi$-plane) slice of the simulation domain at $r/R_\odot=35$.
The four panels show the normalized radial magnetic field ($B_r/B$, top left), radial velocity in unit of ${\rm km \ s^{-1}}$ (top right), normalized fluctuation in radial Poynting flux ($\delta S_r / \overline{S_r} = S_r/\overline{S_r}-1$, bottom left), normalized fluctuation in density ($\delta \rho / \overline{\rho} = \rho/\overline{\rho}-1$, bottom right), respectively.
The white or black solid line corresponds to the boundary of switchbacks: $B_r = 0$.
\added{The animation of this figure runs from simulation time, $t=100$ to $t=200$ minutes, where $t=0$ is when the system reaches a quasi-steady state.} \\
(An animation of this figure is available.)
}
\label{fig:rlabel69_snapshot_for_paper}
\vspace{1.5em}
\end{figure*}

\subsection{Observational properties}

Before presenting our analysis of the switchbacks in our simulations, we first summarize the main observational characteristics of the magnetic switchbacks observed by PSP:
\begin{enumerate}
    \item Switchbacks are Alfv\'enic in the sense that the jump in the magnetic field $\Delta \vec{B}$ is associated with a jump in the velocity~$\Delta \vec{v}$ given by $\pm \Delta \vec{B}/ \sqrt{4\pi \rho}$, where $\rho$ is the plasma density, and the $+$/$-$ sign corresponds to Alfv\'en waves propagating anti-parallel/parallel to the background magnetic field in the frame co-moving with the bulk solar wind.
    The particular value of this +/- sign for each switchback is opposite to the sign of $B_r$, so that switchbacks always propagate away from the Sun in the plasma frame.
    As a consequence, the switchbacks are associated with enhanced values of the radial component of the plasma velocity \citep{Kaspe19,Bale019,Horbu20}.     
    We note that switchbacks are sometimes observed to have a compressive component \citep{Bale019,Farre20}.
    \item Switchbacks are approximately ``spherically polarized,'' in the sense that the strength of the magnetic field is almost constant as the field abruptly rotates \citep{Bale019,Farre20}.
    \item Switchbacks appear to be field-aligned, elongated structures with a typical aspect ratio of $10$ near the first perihelion \citep{Horbu20,Laker20}, where the aspect ratio is defined as the ratio of the length scales parallel and perpendicular to the magnetic field.
    \item A waiting-time analysis  indicates that switchbacks appear in clusters \citep{Dudok20}.
    As a result, although the filling factor of $B_r$-reversed regions is $6 \%$ during PSP's first perihelion encounter \citep{Bale019}, the ``active phase'' of the solar wind that contains switchbacks  occupies at least $75 \%$ of the total observation time near the first perihelion \citep{Horbu20}.
    \item Switchbacks are S-shaped, locally folded magnetic-field lines.
    The propagation directions of the electron strahl \citep{Kaspe19,Whitt20},  small-scale Alfv\'en waves \citep{McMan20},
    and proton-alpha differential velocity \citep{Yamau04b} support this picture.
    \item The proton core parallel temperature, the field-aligned temperature obtained from the core of the proton distribution function, is found to be constant
    across switchbacks \citep{Wooll20}, 
    possibly because switchbacks  move through the plasma at the local Alfv\'en speed, which prevents them from strongly heating any particular parcel of plasma before propagating past it.
    We note, however, that the active phase mentioned above is characterized by an enhanced proton parallel temperature \citep{Woodh20}.
    \item The volume filling factor of switchbacks is observed to increase
as the heliospheric distance increases from $\sim35 R_{\odot}$ to $\sim50 R_{\odot}$ \citep{Mozer20}.
In light of the comparative rarity of switchbacks near $1 {\rm \ au}$, the volume filling factor should begin to decrease somewhere between $50 R_\odot$ and $1 {\rm \ au}$.
\end{enumerate}
The theory of magnetic switchbacks should explain these observational properties.

\begin{figure}[t]
\centering
\vspace{-1em}
\includegraphics[width=72mm]{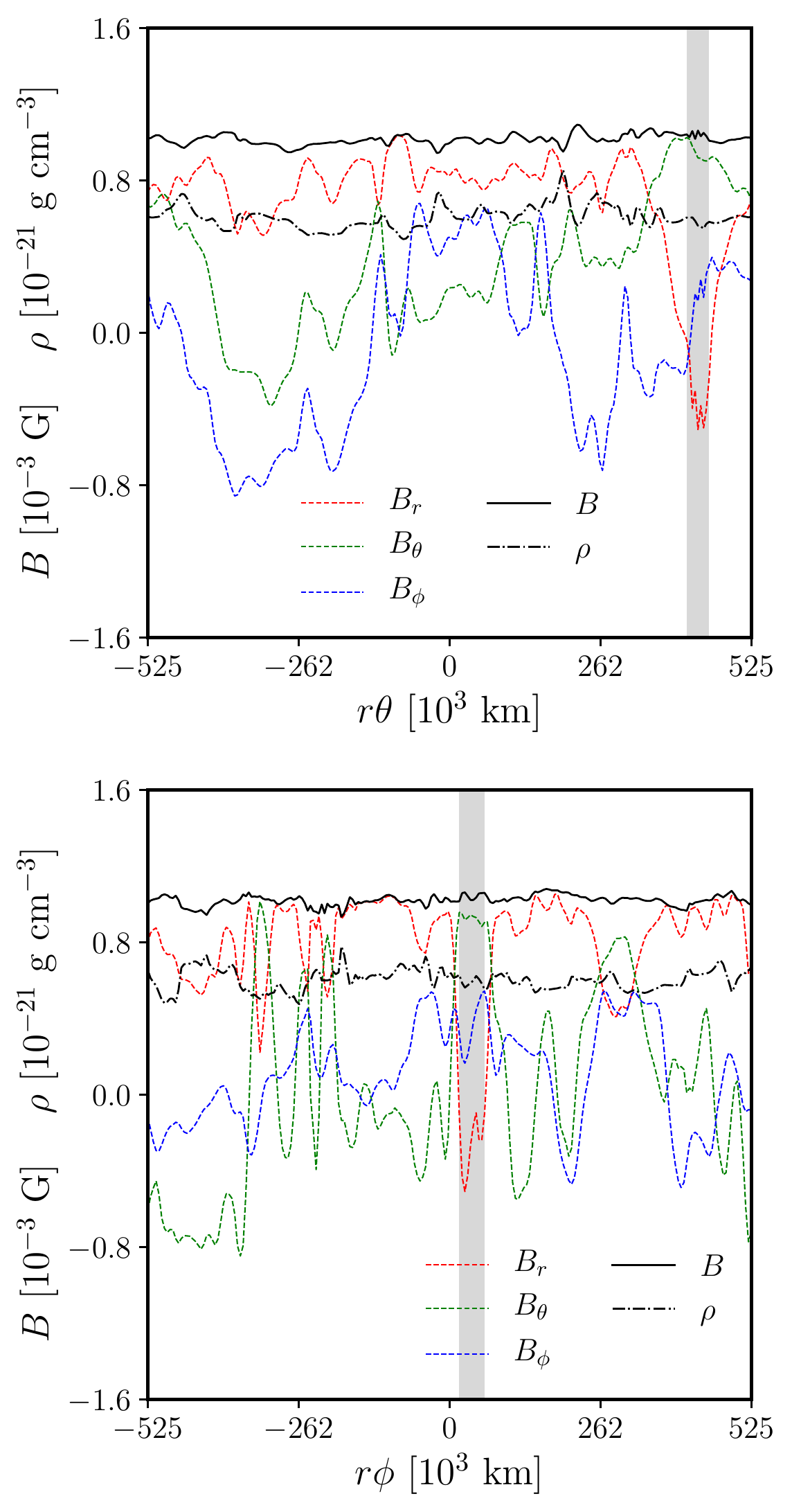}
\caption{
An example of the one-dimensional structure of a magnetic switchback at $r/R_\odot = 35$.
The profiles along the dashed and dotted lines in Figure \ref{fig:rlabel69_snapshot_for_paper} are displayed in the top and bottom panels, respectively, where the switchback region is highlighted by the shaded rectangle. 
}
\vspace{1em}
\label{fig:switchback_1D_profile}
\end{figure}

\begin{figure}[t]
\centering
\vspace{-1em}
\includegraphics[width=80mm]{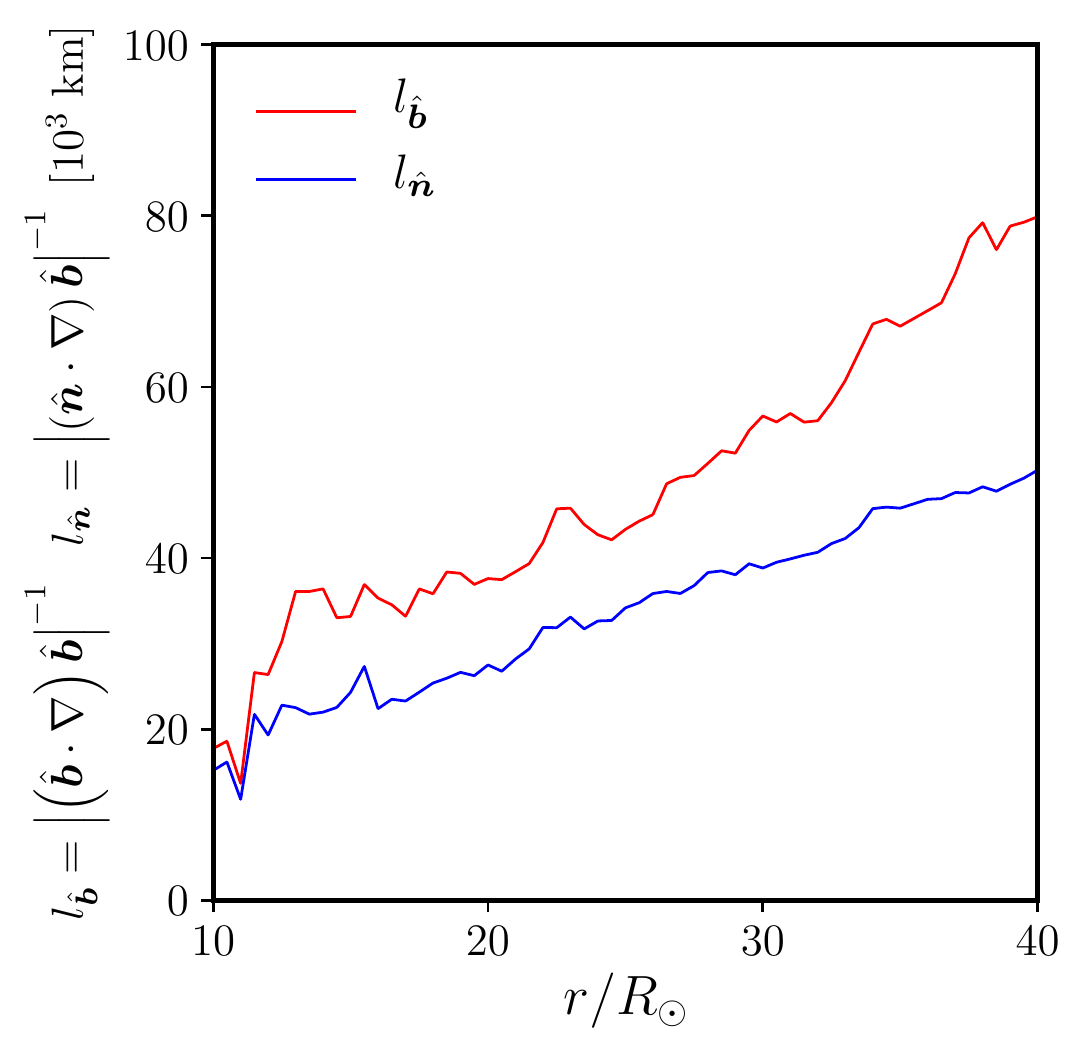}
\caption{
Radial variation of statistically averaged $l_{\hat{\vec{b}}}$ (red line) and $l_{\hat{\vec{n}}}$ (blue line).
}
\vspace{1em}
\label{fig:switchback_length_scale}
\end{figure}

\subsection{Structure in the $\theta\phi$ plane \label{sec:horizontal_structure_sb}}

To see the localized nature of magnetic switchbacks,
we show in Figure \ref{fig:rlabel69_snapshot_for_paper} the simulation data in the  $\theta\phi$ plane at $r/R_\odot = 35$.
The four panels show the normalized radial magnetic field $B_r/B$, density fluctuation $\delta \rho / \overline{\rho} = \rho / \overline{\rho} - 1$, radial Poynting-flux fluctuation $\delta S_r / \overline{S_r} = S_r / \overline{S_r} - 1$, and radial velocity $v_r$.
The white solid lines in the top-right and lower-left panels and the black solid lines in the lower-right panel correspond to the boundaries of magnetic switchbacks (where $B_r = 0$).
Note that the overline denotes an average in the  $\theta\phi$~plane.

\begin{figure*}[t]
\centering
\includegraphics[width=170mm]{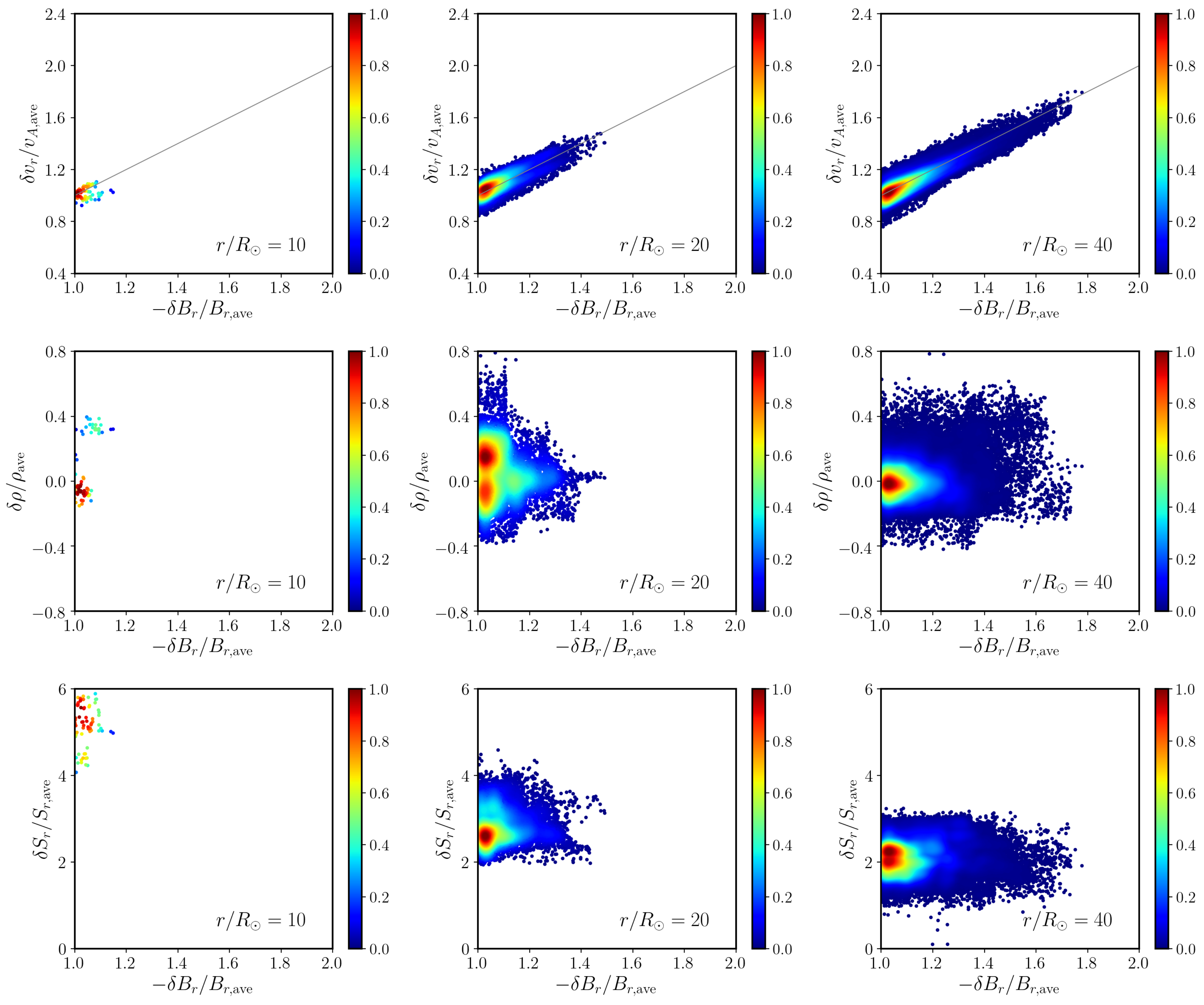}
\caption{
Scatter plots of the normalized radial-field fluctuation $- \delta B_r / B_{r,{\rm ave}}$ and radial-velocity enhancement ($\delta v_r / v_{Ar,{\rm ave}}$, top panels), 
density fluctuation ($\delta \rho / \rho_{\rm ave}$, middle panels), and
Poynting-flux fluctuation ($\delta S_r / S_{r,{\rm ave}}$, bottom panels).
Left, center and right panels correspond to $r/R_\odot=10$, $20$, and $40$, respectively.
The color represents the normalized kernel density estimate of the points.
Grey lines in the top panels show the Alfv\'enic-correlation relation $\delta B_r / B_{r,{\rm ave}} = - \delta v_r / v_{Ar,{\rm ave}}$.
}
\label{fig:switchback_events_correlation}
\vspace{1em}
\end{figure*}

A careful observation of the $\theta\phi$-plane structure reveals several physical properties of the switchbacks:
\begin{enumerate}
    \item As seen in the top-left panel of Figure \ref{fig:rlabel69_snapshot_for_paper}, in which the switchback (blue patch) is much smaller than the $r\theta$ plane,
    the switchbacks are highly-localized structures in that their horizontal spatial extent is much smaller than the energy-containing scale of the turbulence, which is comparable to the size of simulation domain in the $\theta$ and $\phi$ directions.
    \item The switchbacks are always associated with enhancements in radial velocity and radial Poynting flux, 
    as illustrated by the top-right and bottom-left panels.
    Meanwhile, there does not appear to be any correlation between the emergence of switchbacks and density fluctuations, 
    since no clear enhancement in the density fluctuation is seen in the switchback region of the bottom-right panel.
\end{enumerate}
These characteristics are more quantitatively discussed in the following sections.

The structure of a magnetic switchback can be clearly seen in a plot of its one-dimensional spatial profile.
Figure \ref{fig:switchback_1D_profile} shows an example of the one-dimensional structure of a switchback along the white dashed and dotted lines in the left-top panel of Figure \ref{fig:rlabel69_snapshot_for_paper}.
An interesting point is that the total magnetic field is nearly constant across the switchback in both the $\theta$ and $\phi$ directions,
which is inferred from the $B$ profile (black solid line) in Figure \ref{fig:switchback_1D_profile}.
Another interesting behavior (which is especially clear in the bottom panel) is that the rapid decrease in the $B_r$ component is compensated for by the rapid increase in the perpendicular component 
($B_\phi$ in the top panel and $B_\theta$ in the bottom panel) while the parallel component 
($B_\theta$ in the top panel and $B_\phi$ in the bottom panel) is nearly constant.

To investigate whether switchbacks are closer to rotational discontinuities (RD) or tangential discontinuities (TD) \citep[see, e.g.,][]{Horbu01},
we define the field-aligned and field-normal length scales
\citep{Squir20}
\begin{align}
    l_{\hat{\vec{b}}} = \left| \left( \hat{\vec{b}} \cdot \nabla \right) \hat{\vec{b}} \right|^{-1} \ \ \
    l_{\hat{\vec{n}}} = \left| \left( \hat{\vec{n}} \cdot \nabla \right) \hat{\vec{b}} \right|^{-1},
\end{align}
where $\hat{\vec{b}} = \vec{B}/B$ and $\hat{\vec{n}}$ is a unit vector normal to $\hat{\vec{b}}$ ($\hat{\vec{n}} \cdot \hat{\vec{b}} = 0$).
Taking $\hat{\vec{n}}$ to be in the $\theta\phi$ plane, at each given radial distance, $l_{\hat{\vec{b}}}$ and $l_{\hat{\vec{n}}}$ are calculated at the boundaries of switchbacks (where $B_r=0$) and then averaged over at each radial distance. 
To be specific, the total number of detected boundaries is 192 at $r/R_\odot =10$ and 155002 at $r/R_\odot=40$,
and $l_{\hat{\vec{b}}}$ and $l_{{\hat{\vec{n}}}}$ values calculated at these boundary grids are averaged for each $\theta\phi$ plane.

The quantities $l_{\hat{\vec{b}}}$ and $l_{\hat{\vec{n}}}$ represent the length scales of the field deflection in the parallel and perpendicular directions.
If the boundary of a switchback is TD-like, the maximum-variation direction is nearly perpendicular, which yields $l_{\hat{\vec{b}}} > l_{\hat{\vec{n}}}$.
The radial variations of averaged $l_{\hat{\vec{b}}}$ and $l_{\hat{\vec{n}}}$ are shown in 
Figure \ref{fig:switchback_length_scale}.
The field-aligned length scale $l_{\hat{\vec{b}}}$ is found to be statistically larger than the field-normal length scale $l_{\hat{\vec{n}}}$.
We thus conclude that switchbacks are on average more TD-like than RD-like,
which is consistent with the observational result that the discontinuities in the solar wind are mostly TD-like \citep{Horbu01}.

\begin{figure}[t]
\centering
\includegraphics[width=78mm]{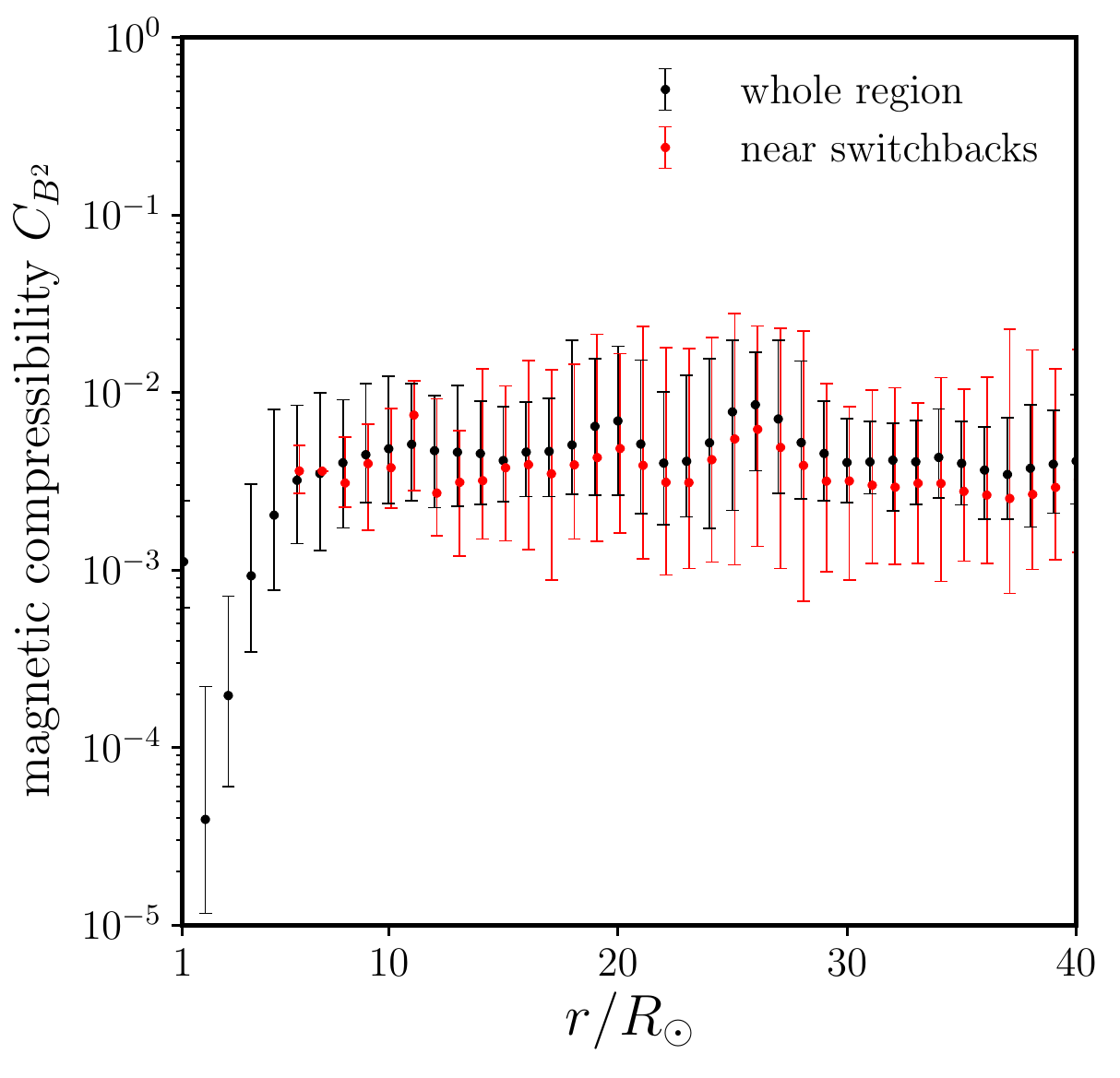}
\caption{
Radial trend of magnetic compressibility $C_{B^2}$.
Upper and lower ends of bar represent the maximum and minimum values and the point indicates the averaged value.
Black bars and points are obtained from $\theta\phi$-plane average while red bars and points are from local average near switchbacks.
}
\label{fig:magnetic_compressibility}
\vspace{1em}
\end{figure}
\subsection{Alfv\'enic nature \label{sec:alfvenic_correlation}}

Important physical properties of magnetic switchbacks can be inferred from the correlations between their
magnetic fluctuations and other quantities.
One of the important correlations is found between the radial-field fluctuation $\delta B_r$ and the radial-velocity fluctuation $\delta v_r$, which are connected by the following Alfv\'enic relation: 
\begin{align}
    - \delta B_r / B_{r,{\rm ave}} \approx \delta v_r / v_{Ar,{\rm ave}},
    \label{eq:alfvenic_correlation}
\end{align}
where $B_{r,{\rm ave}}$ and $v_{Ar,{\rm ave}}$ are the averaged radial magnetic field and radial Alfv\'en speed $v_{Ar} = B_r / \sqrt{4 \pi \rho}$.
In addition to the radial velocity enhancement,
some of the observed magnetic switchbacks are associated with density and Poynting-flux enhancements \citep{Bale019}.
Here we investigate whether these correlations are found in our simulation.

Figure \ref{fig:switchback_events_correlation} shows scatter plots of normalized $B_r$ fluctuations ($-\delta B_r / B_{r,{\rm ave}}$) and normalized radial-velocity enhancements $\delta v_r / v_{Ar,{\rm ave}}$ (top panels),
normalized density fluctuations $\delta \rho / \overline{\rho}$ (middle panels),
and normalized Poynting-flux fluctuations $\delta S_r / \overline{S_r}$ (bottom panels) at different radial distances, 
with the color scale indicating the probability of finding each combination of variables in the simulation domain. 
The radial component of the Poynting flux $S_r$ is defined by
\begin{align}
    S_r = \left( \frac{B_\theta^2}{4 \pi} + \frac{B_\phi^2}{4 \pi} \right) v_r - \frac{B_r}{4 \pi} \left( v_\theta B_\theta + v_\phi B_\phi \right).
\end{align}
The grey lines in the top panels show the Alfv\'enic relation, Eq. (\ref{eq:alfvenic_correlation}). 
Note that the data shown in Figure \ref{fig:switchback_events_correlation} is restricted to switchback regions in which $\delta B_r + B_{r,{\rm ave}}<0$. 
Several features are found in these scatter plots:
\begin{enumerate}
    \item Fluctuations in the radial magnetic field are always associated with fluctuations in the radial velocity.
    All the switchback events approximately satisfy the Alfv\'enic relation, Eq. (\ref{eq:alfvenic_correlation}), 
    indicating the Alfv\'enic nature of magnetic switchbacks.
    \item No clear correlations are found between density fluctuations and switchbacks, in that the majority of switchback events lie near $\delta \rho = 0$, regardless of the switchback amplitude $-\delta B_r / B_{r,{\rm ave}}$.
    \item Magnetic switchbacks always exhibit larger-than-average Poynting flux.
    At the same time, within the switchback population, the switchback amplitude is not correlated with the magnitude of the Poynting-flux fluctuation.
    This means that switchbacks with arbitrary amplitude can emerge once the radial Poynting flux exceeds a critical threshold value that depends on heliocentric distance.
\end{enumerate}
These results indicate that the switchbacks in our simulation are associated with large-amplitude, uni-directional Alfv\'en waves that exhibit strong $v$--$B$ correlation, weak density fluctuations, and larger-than-average radial Poynting flux.

\subsection{Magnetic compressibility \label{sec:magnetic_compressibility_sb}}

The constant-$B$ nature of switchbacks found in Figure \ref{fig:switchback_1D_profile}  is more quantitatively seen from the magnetic compressibility, which is given by
\begin{align}
    C_{B^2} = \left( \frac{\delta \left| \vec{B} \right|}{\left| \delta \vec{B} \right|} \right)^2 = \frac{\overline{\left| B - \overline{B} \right|^2}}{\overline{\left| \vec{B} - \overline{\vec{B}} \right|^2}},
\end{align}
where $B = \left| \vec{B} \right|$ and the overline denotes a spatial average.

Black and red symbols in Figure \ref{fig:magnetic_compressibility} show the radial evolution of magnetic compressibility in the whole $\theta \phi$ plane and near switchbacks (in the square region $[-\theta_{\rm max}/4 \le \theta \le \theta_{\rm max}/4] \times [-\phi_{\rm max}/4 \le \phi \le \phi_{\rm max}/4]$ centered at the local minimum of $B_r$), respectively.
The magnetic compressibility is calculated at each time step and thus has a range of values corresponding to different times.
The upper and lower ends of the bars show the maximum and minimum values of $C_{B^2}$, respectively, and the points represent the mean values. 
The magnetic compressibility defined in the whole $\theta \phi$-plane is much smaller than unity which is consistent with observations in Alfv\'enic solar wind.
Interestingly, the magnetic compressibility near switchbacks is similar to or even smaller than that defined in the whole $\theta \phi$ plane.
In terms of $B$, this analysis shows that the switchbacks are approximately ``spherically polarized,'' in the sense of having nearly constant~$B$, consistent with many of the switchbacks observed by PSP.

Some of the observed switchbacks, on the other hand, have a compressional nature.
If we define ``compressional events" as having $C_{B^2}>0.02$, then $\sim 0.1$--$1 \%$ of switchbacks are compressional, which is much smaller than the observed fraction \citep[$\sim 27\%$, see][]{Laros20}.
Our results suggest that the explanation of such compressional switchbacks lies beyond the mechanisms seen in the simulations presented here and requires additional physics.

\subsection{Aspect ratio}

PSP measurements suggest that switchbacks are elongated along the background magnetic field, with a typical aspect ratio of 10 \citep{Horbu20,Laker20}.
It is worth investigating whether this large aspect ratio is reproduced in our simulation.
We used the following method to measure the aspect ratio of magnetic switchbacks. 
First, at each time step, we find the minimum-$B_r$ grid point in a given $r\theta$ plane. 
When the minimum $B_r$ is negative, we define the minimum-$B_r$ point as a switchback region. 
For each point in a switchback region, the adjacent grid points that satisfy $B_r<0$ are iteratively found and collectively defined as the switchback region. In this way, we obtain a certain connected domain that satisfies $B_r<0$ everywhere. 
The ratio between the radial extent $l_\parallel^{\rm SB}$ and transverse extent $l_\perp^{\rm SB}$ of the domain is defined as the aspect ratio: $l_\parallel^{\rm SB}/l_\perp^{\rm SB}$. 
We note that multiple switchbacks can be found at given $\phi$ and $t$, among which we only focus on the one with minimum $B_r$.

\begin{figure}[t]
\centering
\includegraphics[width=80mm]{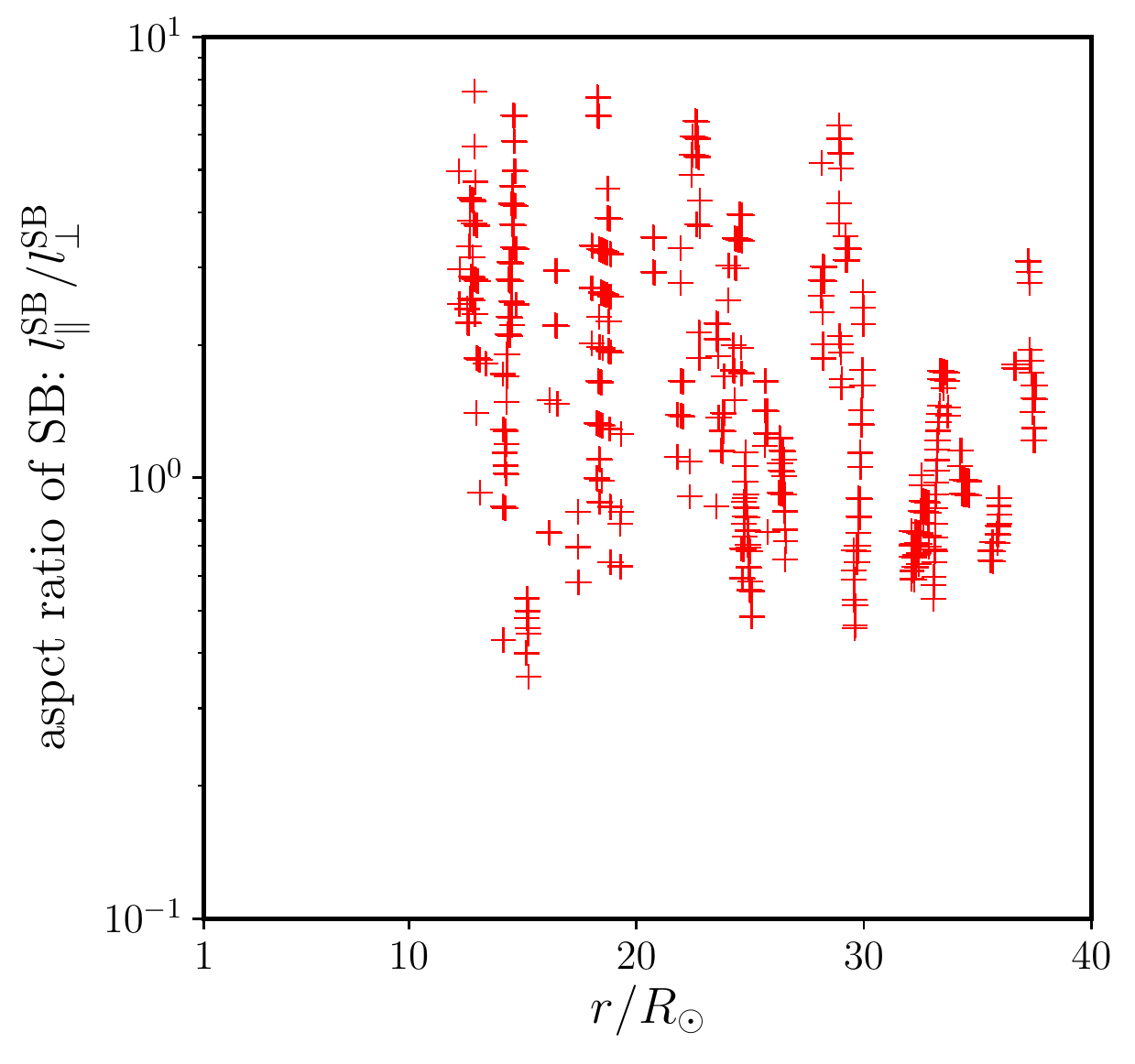}
\caption{
Aspect ratio of the magnetic switchbacks as a function of radial distance. Each point in the figure corresponds to one snapshot.
}
\label{fig:sb_aspect_ratio}
\vspace{1em}
\end{figure}

Figure \ref{fig:sb_aspect_ratio} shows the measured aspect ratio as a function of radial distance.
Although the data are highly scattered, on average the aspect ratio is larger than unity, indicating elongated structure along the mean-field direction ($r$ axis).
The aspect ratio also seems to decrease with $r$.
However, we need to note that the value of aspect ratio is highly influenced by numerical resolution.
Most of the switchbacks are resolved by a few grid points in the perpendicular ($\theta$ and $\phi$) directions and thus are broadened by numerical dissipation.
That the aspect ratio is smaller than the typical observed value \citep[$\sim 10$, see ][]{Horbu20,Laker20}
is possibly due to  insufficient resolution.
The actual value of the aspect ratio should be investigated in future high-resolution simulations.

\begin{figure}[t]
\centering
\includegraphics[width=75mm]{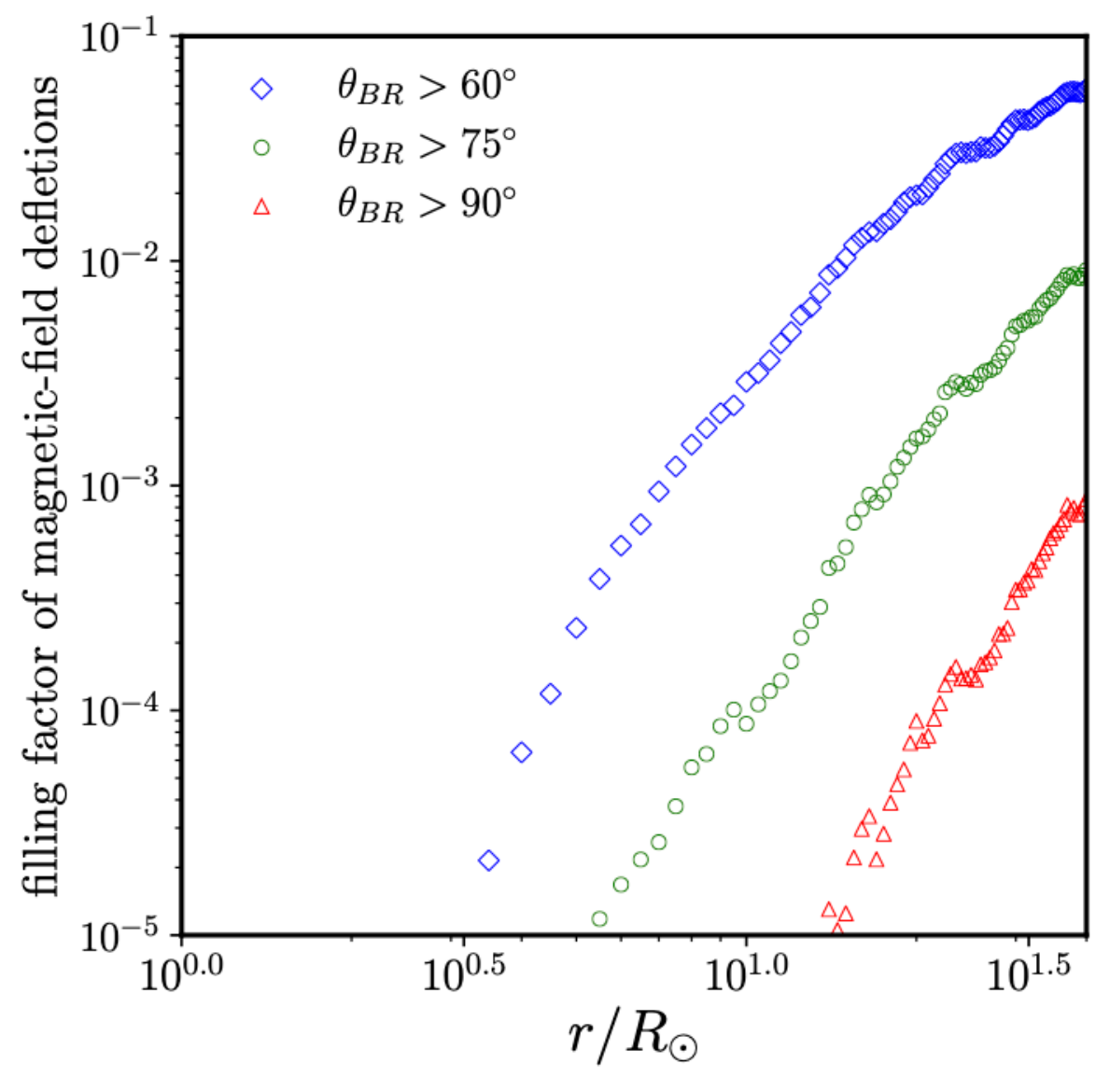}
\caption{
Filling factor of field deflection events versus radial distance.
Blue, green and red symbols correspond to the threshold angles of $\theta_{BR} = 60^\circ$, $\theta_{BR} = 75^\circ$, and $\theta_{BR} = 90^\circ$, respectively. 
}
\label{fig:ff_switchbacks}
\vspace{1em}
\end{figure}

\subsection{Filling factor and amplitude histogram \label{sec:filling_factor_sb}}

The number of field-deflection events, 
the events with large fluctuations in $B_r$ including switchbacks,
is found to increase with distance from the Sun regardless of the deflection angle \citep[see Figure 3 of][]{Mozer20} 
This trend is observed in our simulation data. 
To show this, we define the filling factor of field-deflection events via the following procedure.
We first count the number of grid points that satisfy the criteria  $\theta_{BR}>60^\circ$, $\theta_{BR}>75^\circ$, and $\theta_{BR}>90^\circ$, where $\theta_{BR}$ is the angle between the radial direction and the local magnetic field, in each $\theta\phi$ plane over the 2000 time steps we analyze.
The filling factor of each field-deflection event is then defined as a ratio of the cumulative counts to the total number of $\theta\phi$-plane grid points over 2000 steps ($216 \times 216 \times 2000 = 93,312,000$).
Figure \ref{fig:ff_switchbacks} shows the radial evolution of the filling factors of field-deflection events with $\theta_{BR}>60^\circ$ (blue), $\theta_{BR}>75^\circ$ (green), and $\theta_{BR}>90^\circ$ (red).
The deflection-angle thresholds ($60^\circ$, $75^\circ$ and $90^\circ$) are chosen arbitrarily to investigate if the behavior depends on the threshold angle.
Regardless of the deflection-angle threshold, 
the filling factor of field-deflection events increases with radial distance.
We note, however, that the filling factor of magnetic switchbacks ($\theta_{BR}>90^\circ$) is much smaller than the observed value at $r \approx 35 R_\odot$ \citep[$\sim 6 \%$, see][]{Bale019}.
Specifically, the filling factor in our simulation is $0.063 \%$ at $r/R_\odot = 35.5$ (close to the first perihelion of PSP), and thus,
there exists a two-orders-of-magnitude gap between the observation and the current simulation.

\begin{figure}[t]
\centering
\includegraphics[width=75mm]{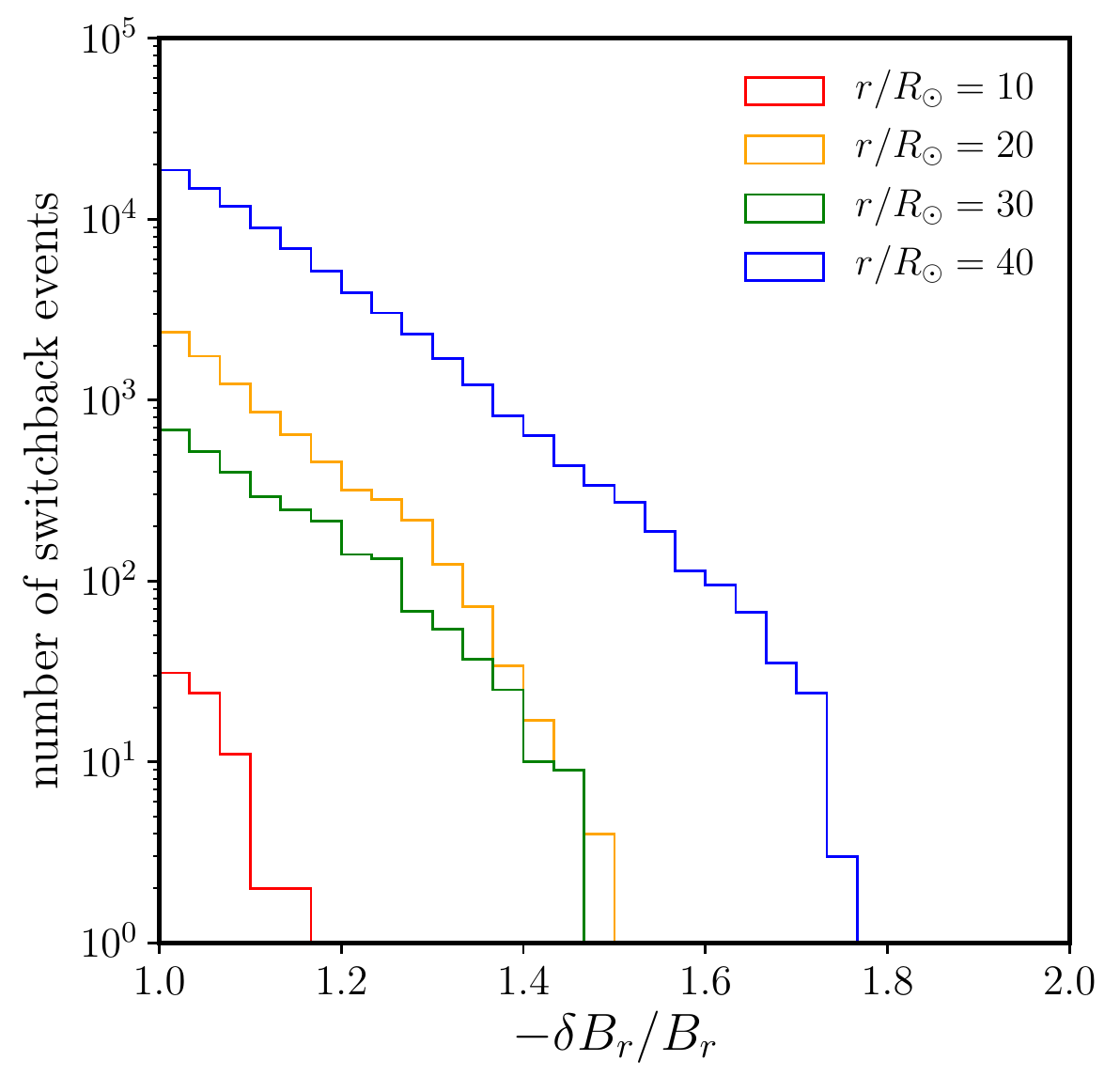}
\caption{
Histogram of normalized switchback amplitudes at four radial distances ($r/R_\odot = 10, 20, 30, 40$).
}
\label{fig:switchback_histogram}
\vspace{1em}
\end{figure}

The smaller filling factor of switchbacks in our simulation may be due to some combination of a smaller turbulence amplitude and insufficient resolution. 
In our simulation, $\left| \delta B_{\rm rms}/B_{r,{\rm ave}} \right| = 0.695$, whereas $\left| \delta B_{\rm rms}/B_{r,{\rm ave}} \right|$ ranged from $0.670$ to $1.280$ in the two intervals from PSP’s first perihelion encounter listed in Table \ref{table:psp_obs_characteristics}. 
As shown in Figure 3 of \citet{Squir20}, the switchback volume filling fraction is highly sensitive to the value of $\left| \delta B_{\rm rms}/B_{r,{\rm ave}} \right|$. 
The presence of intervals during PSP’s first perihelion encounter with larger turbulence amplitudes than our simulation may thus at least partially account for the difference in switchback filling fractions.
In addition, \citet{Squir20} showed that the number of switchbacks is sensitive to the numerical resolution.
In their expanding-box simulations, which are designed to emulate the evolution of Alfv\'enic turbulence out to~$r=35.7 R_{\odot}$, the switchback filling fraction reaches over $ 3\%$ in their highest-resolution run, which has $540\times 1120^2$ grid points. 
In our simulations, if we reduce the resolution of the $\theta \phi$ plane from $(216,216)$ to $(96,96)$, the filling factor of magnetic switchbacks is reduced by a factor of $10$.
Thus, it is possible that the filling factor of magnetic switchbacks would increase substantially in a higher-resolution version of the simulation we have presented.

The distribution of switchback amplitudes is also of interest.
Figure \ref{fig:switchback_histogram} shows histograms of detected switchback events as functions of normalized switchback amplitude $\delta B_r/B_r$.
Here a grid point with negative $B_r$ is counted as a switchback event.
As shown by this histogram, as well as the number of events, the maximum amplitudes of switchbacks are found to increase with radial distance.

\subsection{Propagation speed of magnetic switchbacks \label{sec:propagation_sb}}

Since the simulation box extends globally from the coronal base to $40R_\odot$,
once a magnetic switchback emerges, we are able to measure its propagation velocity by tracing it.
Specifically, we measure how the grid position with minimum $B_r$ moves in time.
Although this rough estimation yields only a limited number of switchback detections, the number of detected events is sufficiently large to discuss their statistical properties.

\begin{figure}[t]
\begin{flushleft}
\includegraphics[width=75mm]{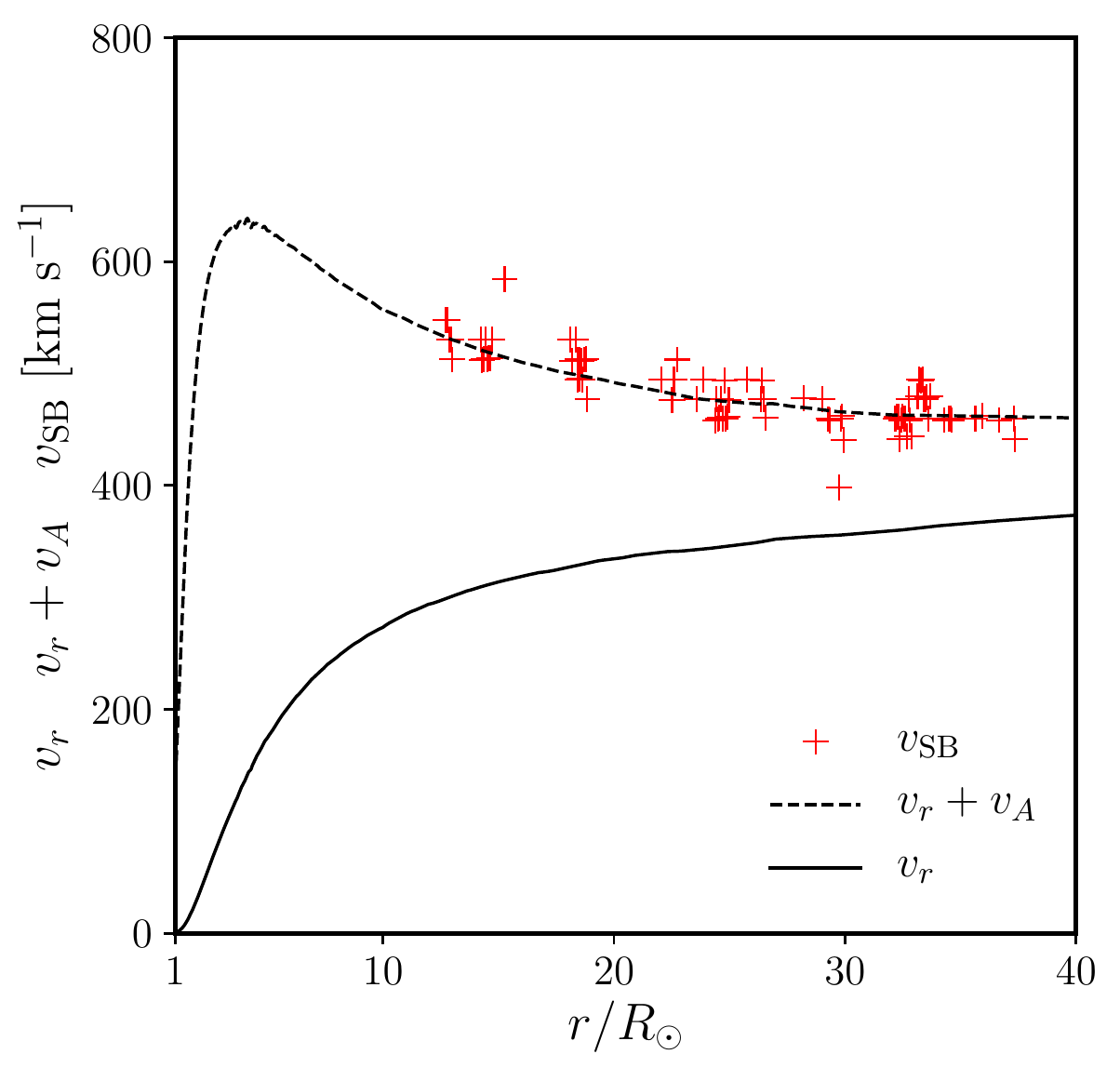}
\end{flushleft}
\caption{Detected propagation speed of magnetic switchbacks ($v_{\rm SB}$, red crosses).
Black lines are averaged radial velocity ($v_{r,{\rm ave}}$, solid line) and radial velocity plus Alfv\'en speed ($v_{r,{\rm ave}} + v_{A,{\rm ave}}$, dashed line).
}
\label{fig:sb_speed}
\vspace{1em}
\end{figure}

Figure \ref{fig:sb_speed} compares the detected switchback velocity $v_{\rm SB}$ with the averaged radial velocity $v_{r,{\rm ave}}$ (black solid line) and radial velocity plus Alfv\'en speed $v_{r,{\rm ave}} + v_{A,{\rm ave}}$ (black dashed line).
Clearly the radial profile of $v_{\rm SB}$ matches $v_{r,{\rm ave}} + v_{A,{\rm ave}}$,
supporting the idea that magnetic switchbacks are locally bent field lines that propagate through the plasma in the anti-Sunward direction at the local Alfv\'en speed.

\section{Summary and discussion}

In this work, we have performed a direct numerical simulation of the wave/turbulence-driven solar wind.
Our simulation yields  Alfv\'enic slow solar wind, the same type of solar wind observed in the first perihelion encounter of PSP.
The radial profiles of the density, velocity, and magnetic field are in agreement with remote-sensing observations and in-situ measurements (Figure \ref{fig:mean_field}).

The turbulence in our simulations is generated by launching outward-propagating Alfv\'en waves into our simulation domain through the inner (coronal-base) boundary. These waves become turbulent as they propagate away from the sun and ultimately form magnetic switchbacks.  A possible scenario for switchback generation inferred from our simulation is summarized as follows.
\begin{enumerate}
    \item The normalized amplitude of the Alfv\'en waves ($\delta B/B$) increases with $r$ due to the expansion of the background flux tube and the decrease of the plasma density 
    \citep[Figure \ref{fig:vb_fluctuation}, see][]{Parke65,Heine80}.
    \item Outward-propagating Alfv\'en waves nonlinearly evolve towards a constant-$B$ or ``spherically polarized'' state 
    \citep[Figure \ref{fig:magnetic_compressibility}, see][]{Cohen74a,Barne74, Vasqu98}.
    \item Once $\delta B/B$ increases to values~$\sim 1$, the ongoing nonlinear drive towards spherical polarization generates discontinuities in the field, because it is (or at least appears to be) impossible to form a 3D state with constant-$B$ and $\delta B/B \sim 1$ \citep[Figure \ref{fig:switchback_1D_profile}, see][]{Barne76,Valen19}.
\end{enumerate}

The switchbacks in the simulation have several of the properties exhibited by observed switchbacks, including spherical polarization (Figures \ref{fig:switchback_1D_profile}, \ref{fig:magnetic_compressibility}), Alfv\'enic $v$--$b$ correlation (Figure \ref{fig:switchback_events_correlation}), and
a volume filling factor that increases with~$r$ (Figure \ref{fig:ff_switchbacks}).
Switchbacks are also found to exhibit a rapid change in field direction (Figure \ref{fig:switchback_1D_profile}), to be more `TD-like' than `RD-like' (Figure \ref{fig:switchback_length_scale}), and to propagate outward at the Alfv\'en speed in the frame of background flow (Figure \ref{fig:sb_speed}).
From the simulation results and the discussion above, 
we conclude that at least some of the magnetic switchbacks seen in the solar wind arise as a natural consequence of Alfv\'en waves and turbulence with sufficiently large amplitude. 
However, the volume filling fraction of switchbacks in our simulation is almost two orders of magnitude smaller than in PSP's first perihelion encounter with the Sun, 
possibly because of insufficient numerical resolution and/or the fact that $\left| \delta B_{\rm rms}/B_{r,{\rm ave}} \right|$ is smaller in our simulation than in the PSP data (see Table \ref{table:psp_obs_characteristics}).
Determining the volume filling fraction in simulations with higher-resolution and different values of $\left| \delta B_{\rm rms}/B_{r,{\rm ave}} \right|$ is an important goal for future research in this area.

In addition to its importance for determining the switchback volume filling factor, running simulations with larger numerical resolution will be important for describing the fine-scale structure of switchbacks, including the local reversal of the cross helicity \citep{McMan20}. Besides increasing the numerical resolution, there are a number of other ways that the simulation presented here could be further improved.
Since an artificial energy source term is used, our model does not accurately capture the thermodynamic properties of switchbacks \citep{Wooll20,Woodh20}.
Field-aligned thermal conduction needs to be considered in the future, ideally incorporating the deviation from the Spitzer-H\"arm heat flux in the weakly collisional, large-Knudsen-number regime \citep{Salem03,Bale013,Versc19}.
Also, kinetic effects are ignored in our MHD treatment.
Considering the scale gap between the typical switchback size and the proton gyroradius, the MHD approximation is likely sufficient for describing the dynamics of switchbacks.
However, the temperature is anisotropic  in the solar wind \citep{Helli06}, which can affect  Alfv\'en-wave dynamics \citep{Squir16,Tener18}.
A fluid approximation with temperature anisotropy \citep{Chand11,Hirab16,Tener18} is a possible future generalization of the current simulation.

The authors thank Marco Velli, Kosuke Namekata, Shinsuke Takasao, Jono Squire, Romain Meyrand, Alfred Mallet, Stuart Bale, Justin Kasper, Tim Horbury, and Gary Zank for
valuable discussions and comments.
The authors are also grateful to the anonymous referee for valuable comments.
We acknowledge the NASA Parker Solar Probe Mission and the FIELDS team led by S. Bale and SWEAP team led by J. Kasper for use of data.
Parker Solar Probe was designed, built, and is now operated by the Johns Hopkins Applied Physics Laboratory as part of NASA’s Living with a Star (LWS) program (contract NNN06AA01C).
Support from the LWS management and technical team has played a critical role in the success of the Parker Solar Probe mission.
Numerical computations were carried out on the Cray XC50 at the Center for Computational Astrophysics, National Astronomical Observatory of Japan. 
MS is supported by a Grant-in-Aid for Japan Society for the Promotion of Science (JSPS) Fellows and by the NINS program for cross-disciplinary study (grant Nos. 01321802 and 01311904) on Turbulence, Transport, and Heating Dynamics in Laboratory and Solar/ Astrophysical Plasmas: “SoLaBo-X.” 
BC is supported in part by
NASA grant
NNN06AA01C to the Parker Solar Probe FIELDS Experiment and by NASA
grants NNX17AI18G and 80NSSC19K0829.

\bibliographystyle{aasjournal}

\end{document}